\documentclass[journal,twocolumn]{IEEEtran}

\usepackage{amsmath,graphicx,algorithmic,float,cite,dsfont,amssymb}
\usepackage{amsfonts,multirow,bm,array,setspace,stfloats}
\usepackage{textcomp}
\usepackage{subfigure}
\usepackage{psfrag}
\usepackage{color}
\usepackage[linesnumbered,ruled]{algorithm2e}
\usepackage{ulem}
\usepackage{pifont}

\ifCLASSINFOpdf
\else
\fi
\newtheorem{proposition}{Proposition}

\newtheorem{theorem}{Theorem}
\newtheorem{corollary}{Corollary}

\newcommand{\bx}{\mathbf{x}}
\newcommand{\sA}{\sqrt{\mathbf A}_m}

\newcommand{\bI}{\bm I}

\begin{document}
%
\title{\color{black}Optimization of MIMO Device-to-Device Networks via Matrix Fractional Programming:\\A Minorization-Maximization Approach}
\author{Kaiming~Shen,~\IEEEmembership{Student Member,~IEEE},
        Wei~Yu,~\IEEEmembership{Fellow,~IEEE},\\
        Licheng~Zhao,
        and Daniel~P.~Palomar,~\IEEEmembership{Fellow,~IEEE}

\thanks{Manuscript submitted to IEEE/ACM Transactions on Networking on \today.
This work is supported in part by Natural Science and Engineering Research Council (NSERC), in part by Huawei Technologies Canada, and in part by the Hong Kong RGC 16208917 research grant and by a Hong Kong Telecom Institute of Information Technology Visiting Fellowship.
The materials in this paper have been presented in part in IEEE International Symposium on Information Theory (ISIT), Aachen, Germany, June 2017 \cite{shen_isit}.}
\thanks{K. Shen and W. Yu are with the Department of Electrical and Computer
Engineering, University of Toronto, Toronto, ON M5S 3G4, Canada (e-mails: kshen@ece.utoronto.ca, weiyu@ece.utoronto.ca).}
\thanks{L. Zhao and D. P. Palomar are with the Department of Electrical and Computer Engineering, Hong Kong University of Science and Technology (HKUST), Clear Water Bay, Kowloon, Hong Kong, China (e-mails: lzhaoai@connect.ust.hk, palomar@ust.hk).
}
}


%


\maketitle


\begin{abstract}
Interference management is a fundamental issue in device-to-device (D2D)
communications whenever the transmitter-and-receiver pairs are located
in close proximity and frequencies are fully reused, so active
links may severely interfere with each other. This paper devises an
optimization strategy named FPLinQ to coordinate the link scheduling decisions
among the interfering links, along with power control and beamforming. The key
enabler is a novel optimization method called matrix fractional programming
(FP) that generalizes previous scalar and vector forms of FP in allowing
multiple data streams per link. From a theoretical perspective, this paper
provides a deeper understanding of FP by showing a connection to the
minorization-maximization (MM) algorithm. From an application perspective, this
paper shows that as compared to the existing methods for coordinating
scheduling in the D2D network, such as FlashLinQ, ITLinQ, and ITLinQ+,
the proposed FPLinQ approach is more general in allowing multiple antennas at
both the transmitters and the receivers, and further in allowing arbitrary and
multiple possible associations between the devices via matching.
Numerical results show that FPLinQ significantly outperforms the previous
state-of-the-art in a typical D2D communication environment.
\end{abstract}



%
\IEEEpeerreviewmaketitle

\section{Introduction}

\IEEEPARstart{S}{pectrum} sharing in an interference-limited wireless
communication environment is one of the most fundamental problems in network
engineering, for which no efficient global optimum algorithm is yet available.
This problem is challenging especially when a large number of mutually
interfering links are present. One essential difficulty lies in deciding which
links should be active at any given time, i.e., how to schedule. But the
optimal scheduling is also intimately related to power control and beamforming,
if the communication links are equipped with multiple antennas, as power and
beam pattern have significant effect on the interference. This coordinated
scheduling, beamforming, and power control problem is important in the emerging
device-to-device (D2D) communication paradigm in which arbitrary peer-to-peer
transmissions can take place, but also relevant in traditional cellular
networks in which coordination among the cells can significant improve the
network performance.

\begin{figure}[t]
\begin{minipage}[b]{0.49\linewidth}
  \centering
	  \centerline{\includegraphics[width=4.2cm]{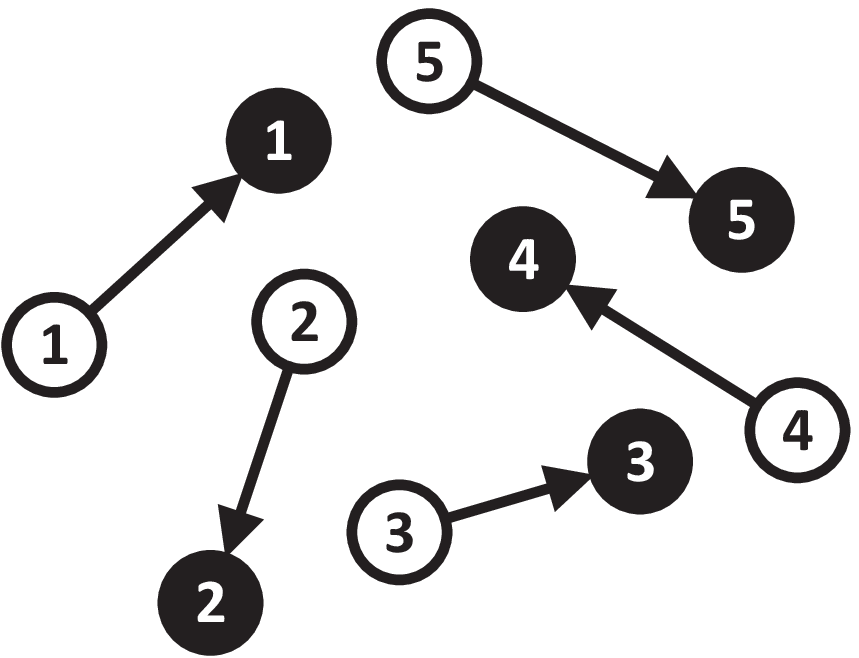}}
  \centerline{(a) \small Fixed Single Association}\medskip
\end{minipage}
\hfill
\begin{minipage}[b]{0.48\linewidth}
  \centering
	  \centerline{\includegraphics[width=4.2cm]{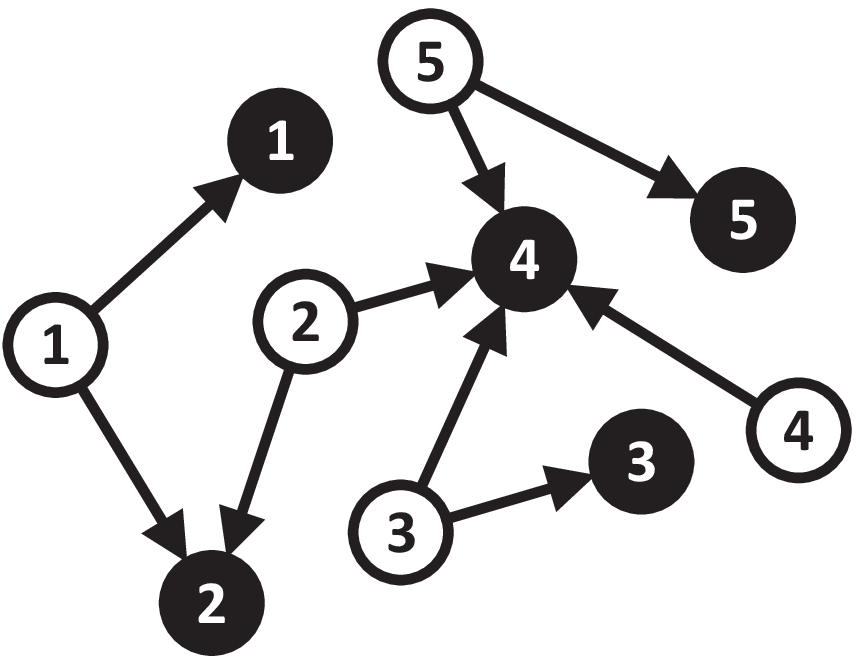}}
  \centerline{(b) \small Flexible Association}\medskip
\end{minipage}
\caption{\color{black}D2D network with white circles denoting the transmitters
and black circles denoting the receivers.  In the fixed single association model (a),
the transmitters have a fixed one-to-one mapping to the receivers.
This paper considers a more general setting (b) in which each
transmitter can be associated with one of multiple receivers, and each receiver
can be associated with one of multiple transmitters. 
}
\label{fig:topology}
\end{figure}

This paper devises a novel optimization technique based on fractional
programming (FP) for solving this classic problem. The problem formulation is
that of maximizing a weighted sum rate of links across a D2D network, in which
the weights account for fairness and the links are selectively activated in
order to alleviate interference.
{\color{black} In addition, this paper considers a model that allows each
transmitter to have the flexibility of associating with one of multiple receivers,
and each receiver to have the flexibility of associating with one of multiple
transmitters,  as illustrated in Fig.~\ref{fig:topology}.}
This overall scheduling, power control, and beamforming problem is a difficult
combinatorial and nonconvex optimization, because the scheduling decision of
each link depends strongly on the activation states and the transmission
parameters (e.g., power and beamforming pattern) of the nearby links.

Motivated by the crucial role of the signal-to-interference-plus-noise
ratio (SINR) in the communication system design, this paper proposes a
\emph{fractional programming based link scheduling} (FPLinQ) strategy to
address the coordinated scheduling, beamforming, and power control problem in an
interference network. While the use of FP for scheduling, beamforming, and power
control originated from our previous work \cite{part1,part2}, the method
proposed in \cite{part1, part2} are restrictive in the sense that: (i) only
scalar and vector cases are treated so that each communication link can only
have one single data stream; (ii) the application regime is restricted to the
cellular setting in which each user is associated with one single fixed base-station.

The present paper significantly generalizes \cite{part1, part2} in several nontrivial directions.
The key theoretical development here is a novel matrix FP technique for dealing with
ratios involving matrices,
in contrast to earlier FP techniques that deal with only the scalar ratio, thus
allowing the full capacity of the multiple-input multiple-output (MIMO) channel
to be realized with multiple data streams. Furthermore, this paper tackles a
more general scheduling problem in which each transmitter/receiver has the
flexibility in associating with each other.
Moreover, this paper makes a theoretical contribution by interpreting the
proposed FP approach as a minorization-maximization (MM) algorithm, thus
allowing convergence to be readily established.  An interesting finding of this
paper is that the FP transforms can be interpreted as novel surrogate functions
in the MM context.

\subsection{Related Work}

Interference-aware scheduling, power control and beaforming for wireless
networks have attracted considerable research interests over the years, e.g.,
\cite{ElBatt_Ephremides_TWC04,Lin_Shroff_TON06,Xi_Yeh_MobiHoc07,Joo_Lin_Shroff_TON09,Huang_JSAC09,Tan_Friedland_Low_JSAC11,Tan_Chiang_Srikant_TON13,Zhuang_Guo_Honig_JSAC15,Zhou_Guo_Mobihoc17}.
In the multiple-antenna cellular network context, the well-known WMMSE
algorithm \cite{luo} is able to attain a stationary point of the joint
power control and beamforming problem; furthermore, under some special
condition for the single-cell case, the global optimum solution can sometime be
found \cite{tan_TSP11}. But the cellular model is different from the model
considered in this paper, because spatial multiplexing can typically be
implemented in a cellular base-station, while the D2D model of this paper only
allows one-to-one mapping between each transmitter and each receiver.

In the D2D context, there is a vast array work in the existing literature
exploring a variety of different directions, including geometry programming
\cite{scale_power}, game theory \cite{game}, stochastic geometry
\cite{stochastic,Lin_Andrews_Ghosh_TWC14}, evolution theory \cite{evolution},
and dynamic programming \cite{DP}. While some of existing works
\cite{ElBatt_Ephremides_TWC04,sheng_QoS,lin_QoS} adopt a quality-of-service
(QoS) model for the scheduling problem, many other works (including this paper)
consider maximizing the weighted sum rate across the D2D network, where the
fairness is taken into account by appropriate setting of the weights.

This paper is most closely related to a series of works that propose algorithms
called FlashLinQ \cite{FlashLinQ}, ITLinQ \cite{ITLinQ}, and ITLinQ+
\cite{ITLinQ+}, which address the D2D scheduling problem using greedy search
while utilizing information theoretic intuition based on generalized
degree-of-freedom (GDoF); we review these algorithms in details in Section
\ref{subsec:otherLinQ}.

An important benchmark method for the problem considered in this paper is the
block coordinate descent (BCD) approach, which is proposed for the cellular network in
\cite{weiyu_TWC}, but can also be adopted for the D2D model. However, BCD is
prone to being trapped in the local optimum solution, as we discussed in
Section \ref{subsec:BCD}.
Using the greedy algorithm and the BCD method as the benchmarks, the aim of
this paper is to show that an optimization motivated approach based on FP can
significantly outperform these state-of-the-art methods.

\subsection{Main Contributions}

The main contributions of this paper are summarized below:
\begin{itemize}
\setlength\itemsep{0.1em}
    \item {\color{black}\emph{Multiple-Antenna Flexible-Association D2D Network Model:}
This work considers a D2D network setup with multiple antennas at both
the transmitters and the receivers, thus each link can carry multiple data
streams. Further, the model considered here allows the flexility among multiple
possible associations between the transmitters and the receivers.
This is a more general model than the ones considered
in the previous works \cite{FlashLinQ,ITLinQ,ITLinQ+,part1,part2}. }

    \item \emph{Matrix FP Transforms:}
	This paper introduces the matrix FP which treats
$\sqrt{\mathbf A}^\dag(\bx) \mathbf B^{-1}(\bx)\sqrt{\mathbf A}(\bx)$ as a ratio between the positive (semi-)definite matrix-valued functions
$\mathbf A(\bx)$ and $\mathbf B(\bx)$,
whereas the previous FP theory focuses on the scalar ratio
${A(\bx)}/{B(\bx)}$ between the real-valued functions $A(\bx)\ge0$ and $B(\bx)>0$
\cite{jorswieck} or the vector case $\mathbf{a}^\dagger(\bx)\mathbf{B}^{-1}(\bx)\mathbf{a}(\bx)$
with the vector function $\mathbf{a}(\bx)$ \cite{part1,part2}.
We extend the FP transforms of \cite{part1,part2} to the matrix case.

	\item \emph{Interpretation of Matrix FP as MM:}
	This paper shows that, from an MM algorithm perspective, the proposed
matrix FP transforms can be thought of as constructing surrogate
functions for the original problem. In this sense, this work puts forward a
novel way of minorizing the logarithmic objective function and the fractional
function, in contrast to the traditional application of MM, which relies on the
second-order Taylor expansion.

    \item \emph{FPLinQ Algorithm:}
This paper proposes an efficient FP based numerical algorithm for the iterative
optimization of scheduling, beamforming, and power control for a D2D network.
It achieves a higher network utility than the previous state-of-the-art.
We observe that the direct optimization of these variables, using for example the
WMMSE algorithm \cite{luo}, may incur a premature link turning-off problem.
In addition, as compared to FlashLinQ, ITLinQ, and ITLinQ+ \cite{FlashLinQ,
ITLinQ, ITLinQ+}, we point out that the information theoretic justification for
ITLinQ actually does not guarantee the optimality of scheduling, and further
the proposed FPLinQ strategy also has an advantage in that it does not require
any tuning of design parameters. {\color{black} The proposed FPLinQ strategy is
based on the decoupling a matrix ratio. There are in fact multiple different
possible decoupling strategies, but the one adopted for FPLinQ is best suited
for algorithmic implementation.}
\end{itemize}

\subsection{Paper Organization and Notations}

The rest of this paper is organized as follows. Section \ref{sec:prob} states
the problem formulation for the wireless joint link scheduling, beamforming,
and power control problem. Section \ref{sec:FP} introduces the matrix FP. We
provide two useful transforms and also connect them to the MM algorithm.
Section \ref{sec:approach} derives the proposed FPLinQ algorithm, Section
\ref{sec:simulation} provides numerical results to validate the performance of
the proposed algorithm. Finally, Section \ref{sec:conclusion} concludes the
paper.

We use lower case, e.g., $s$, to denote scalars, bold lower
case, e.g., $\mathbf{x}$, to denote vectors, and bold upper case, e.g.,
$\mathbf{V}$, to denote matrices. We use $\mathbb R$ to denote the set of real numbers,
$\mathbb C$ to denote the set of complex numbers, {\color{black} $\mathbb H_{++}$ (or $\mathbb H_{+}$) to denote the set of
Hermitian positive definite (or semi-definite) matrices,} $\bI$ to
denote the identity matrix, and $\mathbf 0$ to denote the zero matrix.
We use $(\cdot)^\dagger$ to denote matrix conjugate transpose, {\color{black}$\Re\{\cdot\}$ to denote the real part of a complex number,} and tr$(\cdot)$ to denote matrix trace. We use underline to denote a collection of variables, e.g.,
$\underline{\bx}=\{\bx_1,\bx_2,\ldots,\bx_n\}$ and $\underline{\mathbf
Y}=\{\mathbf Y_1,\mathbf Y_2,\ldots,\mathbf Y_n\}$.

\section{Joint Scheduling, Beamforming, and Power Control in D2D Network}
\label{sec:prob}


\subsection{Problem Formulation}

Consider a wireless D2D network with a set of transmitters $\mathcal I$ and a
set of receivers $\mathcal J$. We assume that each transmitter may have data
to transmit to one or more receivers, and likewise each receiver may wish to
receive data from one or more transmitters. Thus, the communication scenario
considered in this paper is a generalization of traditional D2D network with
fixed single association between each pair of transmitter and receiver
to a scenario with multiple possible associations between the transmitters and
the receivers as shown in Fig.~\ref{fig:topology}.  We assume that in each
scheduling time slot, each transmitter (or receiver) can only communicate with
at most one of its associated receivers (or transmitters)\footnote{Note that
the D2D model considered in this paper is more general than the cellular
network model (e.g., \cite{part2}) in allowing multiple and arbitrary
associations between the transmitters and the receivers, but on the other hand,
is also more restrictive in that it does not allow spatial multiplex at either
the receiver or the transmitter.}, respectively, {\color{black} so that the
mappings between the transmitters and the receivers are one-to-one.} The task
of scheduling is to identify which set of links over the network to activate in
each slot.
Further, we assume that the transmitters and the receivers are each equipped with $N$
antennas and permit multiple data streams to be carried via multiple-input
multiple-output (MIMO) transmission. The task of beamforming and power control
is to design the transmit beamformers for each of these multiple data
streams in each active link in the scheduling slot.

Mathematically, let $\mathcal K_j\subseteq \mathcal I$ be the set of
transmitters associated with each particular receiver $j\in\mathcal J$.
Likewise, let $\mathcal L_i\subseteq \mathcal J$ be the set of
receivers associated with each transmitter $i\in\mathcal I$.
Let $\mathbf{H}_{ji}\in\mathbb C^{N \times N}$ be the channel from transmitter
$i$ to receiver $j$ in the scheduling slot. The joint scheduling, beamforming,
and power control problem can be written down as that of optimizing two sets of
variables: $s_j$, the index of the transmitter scheduled at receiver $j$, and
$\mathbf{V}_i\in\mathbb C^{N\times N}$, the collection of beamforming vectors
at transmitter $i$ in each scheduling slot so as to maximize some network wide
objective function.
Throughout this paper, we assume that the channel state information is
completely known and network optimization is performed in a centralized manner.
We note that this network optimization problem is NP-hard, even under such
idealized assumptions \cite{Sharma_Mazumdar_Shroff_MobiCom06,luo_zhang}.

This paper uses the weighted sum rate as the optimization objective in each
scheduling slot, where the weights are adjusted from slot to slot in an outer
loop in order to maximize a network utility of long-term average rates.
{\color{black}
We assume that interference is treated as noise, so that the achievable data
rate in each scheduling slot can be computed from the receiver's perspective,
i.e., for each receiver $j$, as \cite{goldsmith_book} }
\begin{equation}
\label{rate}
    R_j = \log\Big|\bI+\mathbf{V}^\dag_{s_j}\mathbf{H}^\dag_{js_j}\mathbf F^{-1}_{j}\mathbf{H}_{js_j}\mathbf{V}_{s_j}\Big|
\end{equation}
with the interference-plus-noise term
\begin{equation}
\mathbf F_j = \sigma^2\bI+\sum_{j'\in\mathcal J\backslash\{j\}}\mathbf{H}_{js_{j'}}\mathbf{V}_{s_{j'}}\mathbf{V}^\dag_{s_{j'}}\mathbf{H}^\dag_{js_{j'}}
\end{equation}
where $\sigma^2$ is the power of thermal noise. Given a set of nonnegative
weights $w_{ji}\ge0$, the optimization problem is therefore 
\begin{subequations}
\label{prob:d2d}
\begin{eqnarray}
\underset{\underline{\mathbf{V}},\,\underline{s}}{\text{maximize}} & &
\sum_{j\in\mathcal J}w_{j s_j}R_j
	\label{prob:obj}\\
\text{subject to}
&& \text{tr}\big(\mathbf V^\dag_i\mathbf V_i\big) \le P_\text{max},\;\forall i\in\mathcal I,
    \label{prob:V}\\
&& s_j\in\mathcal K_j\cup\{\varnothing\},\;\forall j\in\mathcal J,
    \label{prob:s_in}\\
&& s_j\ne s_{j'}\text{ or } s_j=\varnothing,\;\forall j\ne j',
    \label{prob:s_ne}
\end{eqnarray}
\end{subequations}
where we have assumed a per-scheduling-slot and per-node power constraint $P_{\max}$
and $\varnothing$ denotes the decision of not scheduling any transmitter at
receiver $j$. We remark that $\mathbf H_{js_j}$, $\mathbf V_{s_j}$, and
$w_{js_j}$ are set to zero if $s_j=\varnothing$. Constraint (\ref{prob:s_ne})
states that the same transmitter cannot be scheduled for more than one receiver
at a time, as required by the assumption that the association between the
transmitters and the receivers in the D2D network must be one-to-one.
Problem (\ref{prob:d2d}) involves a discrete optimization over $\underline s$
and a nonconvex continuous optimization over $\underline{\mathbf V}$, which
make it a challenging optimization problem. Below, we first review several
conventional approaches including the block coordinate descent (BCD) algorithm
and the greedy algorithms.





\subsection{Block Coordinate Descent}
\label{subsec:BCD}


The joint scheduling, beamforming, and power allocation problem as formulated in
(\ref{prob:d2d}) is a mixed discrete-continuous programming problem.  To reach
a reasonable solution, we could optimize the discrete scheduling variable
$\underline s$ and the continuous beamforming variable $\underline{\mathbf V}$
separately and alternatively in a form of the BCD algorithm \cite{weiyu_TWC}. When $\underline s$
is held fixed, optimizing $\underline{\mathbf V}$ alone in (\ref{prob:d2d}) is
the conventional beamforming problem for which existing methods (e.g., the WMMSE
algorithm \cite{luo}) can be applied. When $\underline{\mathbf V}$ is held
fixed, optimizing $\underline s$ alone in (\ref{prob:d2d}) can be recognized as
a weighted bipartite matching problem which can be solved by standard methods.

However, we point out that the BCD approach is prone to a potentially serious
\emph{premature turning-off} problem. Suppose that none of the links related to
a particular transmitter $i$ is scheduled at the early stage of the iterative
optimization, then the beamforming variable $\mathbf V_i$ would be set to
zero. As a result, when $\underline s$ is optimized via matching for the fixed
$\underline{\mathbf V}$ in the next iteration, the matching weights related to
$i$ would all be equal to zero, so the corresponding links can never be turned
back on. Therefore, premature scheduling decisions can adversely affect the overall
performance of the algorithm.


\subsection{FlashLinQ, ITLinQ, and ITLinQ+}
\label{subsec:otherLinQ}

We further examine the current state-of-the-art methods for D2D link scheduling
in the literature: FlashLinQ \cite{FlashLinQ}, ITLinQ \cite{ITLinQ}, and
ITLinQ+ \cite{ITLinQ+}.  These works assume that each terminal has a single
antenna, and further that each transmitter (or receiver) is only associated with one
receiver (or transmitter) respectively, namely the fixed single association case
shown in Fig.~\ref{fig:topology}(a).

Because deciding the \textsc{on-off} state for all the links at the same time
is difficult, all three algorithms propose to schedule the links in a greedy
fashion sequentially, as stated in Algorithm \ref{alg:sequential}. The main
difference between FlashLinQ \cite{FlashLinQ}, ITLinQ \cite{ITLinQ}, and
ITLinQ+ \cite{ITLinQ+} lies in the criterion for deciding whether the new
link conflicts with already scheduled ones in Step 3 of Algorithm \ref{alg:sequential}.

\subsubsection{FlashLinQ \cite{FlashLinQ}}
\label{subsec:flash}
The FlashLinQ scheme \cite{FlashLinQ} applies a threshold $\theta$ to SINR, assuming that adding link $i$ to $\mathcal A$ does not cause conflict if and only if all the activated links have their SINRs higher than $\theta$. The performance of FlashLinQ is highly sensitive to the value of $\theta$, but choosing $\theta$ properly can be difficult in practice. Further, using the same $\theta$ for all the links is usually suboptimal when the weight varies from link to link.

\subsubsection{ITLinQ \cite{ITLinQ} and ITLinQ+ \cite{ITLinQ+}}
\label{subsec:tin}

From an information theory perspective, a seminal study \cite{geng} on the
multi-user Gaussian interference channel provides a sufficient (albeit not
necessary) condition for the optimality of treating interference as noise (TIN)
for maximizing the GDoF as follows:
\begin{equation}
    \label{tin}
    \log |h_{ji}|\ge \max_{j'\ne j}\left\{\log|h_{j'i}|\right\} + \max_{i'\ne i}\left\{\log|h_{ji'}|\right\}
\end{equation}
where $h_{ji}\in\mathbb C$ is the channel of the single-antenna case. We refer to this result as the TIN condition.

\begin{algorithm}[t]
 Initialize the set of activated links $\mathcal{A}$ to $\emptyset$ \;
    \For{each link $(i,j)$}{
        \If{$(i,j)$ does not ``conflict'' with any link in $\mathcal{A}$}{
        Schedule link $(i,j)$ and add it to $\mathcal A$\;
        }
    }
 \caption{Sequential Link Selection}
 \label{alg:sequential}
\end{algorithm}

The central idea of ITLinQ and ITLinQ+ is to schedule a subset of links that
meet this TIN condition. Because the TIN condition in (\ref{tin}) is often too
stringent to activate sufficient number of links, ITLinQ and ITLinQ+ both introduce
relaxation based on design parameters. Like FlashLinQ, the performance of
ITLinQ and ITLinQ+ is heavily dependent on the choice of design parameters,
but they are difficult to choose optimally in practice. For example, \cite{ITLinQ+}
adopts two different sets of parameters for ITLinQ+ for two different network
models. It is often not clear how to adapt ITLinQ and ITLinQ+ to the
particular network environment of interest.

It is important to point out that the theoretical basis of ITLinQ and ITLinQ+,
i.e., the TIN condition, only helps decide whether for some particular
schedule, treating interference as noise is the optimal coding strategy
from a GDoF perspective.
It does not, however, guarantee that if some schedule satisfies the TIN
condition, then it must be the GDoF optimal schedule.
Thus, for a particular network, a schedule that does not satisfy the TIN
condition can outperform one that does.

\begin{figure}
\psfrag{a}{$P$}
\psfrag{b}[][rb]{$P^{0.6}$}
\centering
\centerline{\includegraphics[width=3.5cm]{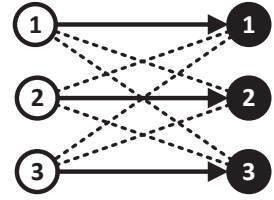}}
\caption{
Power strength is $P$ for each solid signal and is $P^{0.6}$ for each dashed signal. Thus, the sum GDoF equals to 1 if only one link is on, and equals to 1.2 if all links are on.
}
\label{fig:TIN_example}
\end{figure}

This subtle point is illustrated
in the three-link D2D network example shown in Fig.~\ref{fig:TIN_example}.
Let the desired signal strength be $P$ and interfering signal strength be
$P^{0.6}$. At most one link can be activated according to (\ref{tin}), so under the TIN condition, the total GDoF $\le1$.  But, a higher
sum GDoF of 1.2 can be achieved by simply activating all the links.
Therefore, the TIN condition does not guarantee even the GDoF optimality of a
given schedule. Considering further the significant gap between GDoF and the
actual achievable rate, ITLinQ and ITLinQ+ can often produce quite suboptimal
solutions.

In contrast to FlashLinQ, ITLinQ, and ITLinQ+, this paper takes an optimization
theoretic approach of recognizing the optimization objective function as a matrix FP
(since it involves a matrix ratio inside a logarithm), then proposes an iterative
method via a series of matrix FP transforms. The proposed iterative method involves the
update of all the scheduling variables in one step at the same time and the
optimization of beamforming and power variables in subsequent steps. One advantage
of this optimization motivated approach as compared to FlashLinQ, ITLinkQ and
ITLinQ+ is that it does not require the tuning of any threshold parameters.



\section{Matrix Fractional Programming}
\label{sec:FP}

To develop the matrix FP transform, we first present the scalar case as
proposed in \cite{part1,part2}, then provide the matrix generalization. 

\subsection{Scalar FP Transforms}

For scalar FP, the following transform is proposed in \cite{part1} in order to
facilitate optimization by decoupling the numerators and denominators of the scalar
fraction terms in the objective function of an FP.

\begin{proposition}[Quadratic Transform \cite{part1}]
\label{theorem:quadratic}
Given a nonempty constraint set $\mathcal X$ as well as a sequence of nonnegative functions $A_m(\bx)\ge0$, strictly positive functions $B_m(\bx)>0$, and monotonically nondecreasing functions $f_m(z):\mathbb R\mapsto\mathbb R$, for $m=1,2,\ldots,M$, the \emph{sum-of-functions-of-ratio} problem
\begin{subequations}
\label{prob:general_multiple_ratio}
\begin{eqnarray}
\underset{\bx}{{\text{maximize}}} &&
\sum^M_{m=1}f_m\bigg(\frac{A_m(\bx)}{B_m(\bx)}\bigg)
  \label{}\\
{\text{subject to}}&& \bx\in\mathcal{X}
\end{eqnarray}
\end{subequations}
is equivalent to
\begin{subequations}
\label{}
\begin{align}
\underset{\bx,\,\underline{y}}{\text{maximize}}
&\quad\sum^M_{m=1}f_m\left(2y_m\sqrt{A_m(\bx)}-y^2_mB_m(\bx)\right)
	\label{}\\
{\text{subject to}}&\quad \bx\in\mathcal{X}
\end{align}
\end{subequations}
where $y_m\in\mathbb R$ is an auxiliary variable introduced for each ratio term $A_m(\bx)/B_m(\bx)$.
\end{proposition}

The above transform is called the quadratic transform, because it involves
a quadratic function of the auxiliary variables. The quadratic transform
decouples the numerator and the denominator of the fraction, thereby enabling
the iterative optimization between $\mathbf x$ and $y_m$ after the transformation.
This strategy works well for a variety of optimization problems, including
scalar version of the scheduling, beamforming, and power control problem.

Although not immediately recognized at the time \cite{part1} was published, the
above quadratic transform (at least for the case where the functions are trivial, i.e.,
$f_m(z)=z$ for each $m$) is nearly identical to the earlier work of
Benson \cite{Benson1, Benson2}, as restated below.


\begin{proposition}[Benson's Transform \cite{Benson1,Benson2}]
\label{theorem:benson}
Given a nonempty constraint set $\mathcal X$ as well as a sequence of nonnegative functions $A_m(\bx)\ge0$ and strictly positive functions $B_m(\bx)>0$, $m=1,2,\ldots,M$, the \emph{sum-of-ratios} problem
\begin{subequations}
\label{prob:general_multiple_ratio}
\begin{eqnarray}
\underset{\bx}{{\text{maximize}}} &&
\sum^M_{m=1}\frac{A_m(\bx)}{B_m(\bx)}
  \label{}\\
{\text{subject to}}&& \bx\in\mathcal{X}
\end{eqnarray}
\end{subequations}
is equivalent to
\begin{subequations}
\label{}
\begin{align}
\underset{\bx,\underline{u},\underline{v}}{\text{maximize}}&\quad\sum^M_{m=1}\left(2u_m\sqrt{A_m(\bx)}-v_mB_m(\bx)\right)
	\label{}\\
{\text{subject to}}&\quad \bx\in\mathcal{X},\\
&\quad u^2_m-v_m\le0,\;\forall m=1,2,\ldots,M,
\end{align}
\end{subequations}
where $u_m\in\mathbb R$ and $v_m\in\mathbb R$ are introduced as the auxiliary variables for each ratio term $A_m(\bx)/B_m(\bx)$.
\end{proposition}

The above transform is proposed by Benson \cite{Benson1,Benson2} in order to facilitate a branch-and-bound search for the global
optimum of the sum-of-ratios problem. It can be shown that at the optimum, we
must have $u_m^2=v_m$, thus if we had made them equal {\it a priori}, this
reduces to the quadratic transform of Proposition \ref{theorem:quadratic}.

In many practical applications, we wish to optimize functions of ratios.
If the functions are monotonic, then one can apply the quadratic transform
directly as stated in Theorem \ref{theorem:quadratic}. However in case of
logarithmic function, as often encountered in communication system design, a better
alternative is to apply the following transform, proposed in \cite{part2}, to
``move'' the ratio terms to the outside of logarithm. This has an advantage
when discrete (such as scheduling) variables are involved, as it allows
matching algorithms to be used for discrete optimization.


\begin{proposition}[Lagrangian Dual Transform \cite{part2}]
\label{theorem:lagrangian}
Given a nonempty constraint set $\mathcal X$ as well as a sequence of nonnegative functions $A_m(\bx)\ge0$, strictly positive functions $B_m(\bx)>0$, and nonnegative weights $w_m\ge0$, for $m=1,2,\ldots,M$, the \emph{sum-of-logarithmic-ratios} problem
\begin{subequations}
\label{prob:lagrangian}
\begin{eqnarray}
\underset{\bx}{{\text{maximize}}} &&
\sum^M_{m=1}w_m\log\left(1+\frac{A_m(\bx)}{B_m(\bx)}\right)
	\label{}\\
{\text{subject to}}&& \bx\in\mathcal{X}
\end{eqnarray}
\end{subequations}
is equivalent to
\begin{subequations}
\label{prob:lagrangian}
\begin{eqnarray}
\underset{\bx,\,\underline{\gamma}}{{\text{maximize}}} &&
f_r(\bx, \underline{\gamma})
	\label{}\\
{\text{subject to}}&& \bx\in\mathcal{X}
\end{eqnarray}
\end{subequations}
where the new objective function is
\begin{multline}
\label{simplified_fr}
	f_r(\bx, \underline{\gamma}) = \sum^M_{m=1}w_m\log(1+\gamma_m) - \sum^M_{m=1}\gamma_m w_m\\
	+\sum^M_{m=1}\frac{(1+\gamma_m)w_m A_m(\bx)}{A_m(\bx)+B_m(\bx)}
\end{multline}
\end{proposition}
with an auxiliary variable $\gamma_m\in\mathbb R$ introduced for each ratio term $A_m(\bx)/B_m(\bx)$.

The main result of \cite{part2} is that the quadratic transform and the Lagrangian
dual transform can be applied together to decouple the ratio terms in the rate
expression for wireless cellular networks, thus making the network optimization
problem much easier to tackle especially in the presence of discrete scheduling
variables.
{\color{black} To summarize, two different FP techniques are introduced.
Proposition \ref{theorem:quadratic} decouples the numerator and denominator of
the ratio. 
Proposition \ref{theorem:lagrangian} moves the ratio from inside of the
logarithm to the outside.}

The earlier conference version of this paper \cite{shen_isit} further uses the
above transforms for scalar FP to solve the optimal scheduling problem in the
D2D context, but where each transmitter or receiver is equipped with single
antenna.  This paper aims to develop a matrix extension for the MIMO case.

\subsection{Matrix FP Transforms}

The definition of ratio can be naturally generalized to the matrix case. Recall that $\sqrt{\mathbf A}\in\mathbb C^{n\times n}$ is a square root of matrix $\mathbf A\in\mathbb H^{n\times n}_+$ if $\sqrt{\mathbf A}\sqrt{\mathbf A}^\dag=\mathbf A$.
For any pair of $\mathbf A\in\mathbb H^{n\times n}_+$ and $\mathbf B\in\mathbb H^{n\times n}_{++}$, let $\sqrt{\mathbf A}$ be a square root of $\mathbf A$, then $\sqrt{\mathbf A}^\dag\mathbf B^{-1}\sqrt{\mathbf A}$ is said to be a matrix ratio between $\mathbf A$ and $\mathbf B$. The FP transforms of Propositions \ref{theorem:quadratic} and
\ref{theorem:lagrangian} can now be generalized. We state these new
results in the following.

\begin{theorem}[Matrix Quadratic Transform]
\label{theorem:quadratic_matrix}
Given a nonempty constraint set $\mathcal X$ as well as a sequence of numerator functions $\mathbf A_m(\bx)\in\mathbb H_+^{n\times n}$, denominator functions $\mathbf B_m(\bx)\in\mathbb H^{n\times n}_{++}$, and nondecreasing matrix functions $f_m(\mathbf{Z}):\mathbb H^{n\times n}_+\mapsto \mathbb R$ in the sense that $f_m(\mathbf Z')\ge f_m(\mathbf Z)$ if $\mathbf Z'\succeq\mathbf Z$, for $m=1,2,\ldots,M$, the \emph{sum-of-functions-of-matrix-ratio} problem
\begin{subequations}
\label{prob:quadratic_matrix}
\begin{align}
\underset{\bx}{\text{maximize}}\quad&
\sum^M_{m=1}f_m\big(\sqrt{\mathbf A}^\dag_m(\bx)\mathbf B_m^{-1}(\bx)\sqrt{\mathbf A}_m(\bx)\big)
    \label{prob:quadratic_matrix_fo}\\
\text{subject to}\quad& \bx\in\mathcal X
\end{align}
\end{subequations}
is equivalent to
\begin{subequations}
\label{prob:quadratic_matrix:fq}
\begin{align}
\underset{\bx,\underline{\mathbf Y}}{\text{maximize}}\quad&
\tilde f_q(\bx,\underline{\mathbf Y})\\
\text{subject to}\quad& \bx\in\mathcal X,\\
&\mathbf Y_m\in\mathbb C^{n\times n},
\end{align}
\end{subequations}
where the new objective function is defined as
\begin{multline}
\label{quadratic_matrix:fq}
\tilde{f}_q(\bx,\underline{\mathbf Y}) =\\
\sum^M_{m=1}f_m\Big(2\Re\{\sqrt{\mathbf A}^\dag_m(\bx)\mathbf Y_m\}
 - \mathbf Y^\dag_m\mathbf B_m(\bx)\mathbf Y_m \Big).
\end{multline}
Note that (\ref{quadratic_matrix:fq}) implicitly requires that the argument of $f_m(\cdot)$ in (\ref{quadratic_matrix:fq}) is a positive semi-definite matrix.
\end{theorem}
\begin{IEEEproof}
To show that (\ref{prob:quadratic_matrix}) is equivalent to
(\ref{prob:quadratic_matrix:fq}), we first optimize over ${\mathbf Y}_m$ for
fixed $\bx$ in
(\ref{prob:quadratic_matrix:fq}). This can be done for each term in the
summation in $\tilde f_q$ separately. Since $f_m(\cdot)$ is assumed to be monotonic, we only
need to optimize its argument, which is a quadratic function of ${\mathbf Y}_m$.
This optimization has a closed-form solution by completing the square, i.e.,
\begin{align}
&2\Re\{\sqrt{\mathbf A}^\dag_m(\bx)\mathbf Y_m\}
 - \mathbf Y^\dag_m\mathbf B_m(\bx)\mathbf Y_m\notag\\
&=\sqrt{\mathbf A}^\dag_m(\bx)\mathbf Y_m+\mathbf Y^\dag_m\sqrt{\mathbf A}_m(\bx)-\mathbf Y^\dag_m\mathbf B_m(\bx)\mathbf Y_m\notag\\
&= \sqrt{\mathbf A}^\dag_m(\bx)\mathbf B_m^{-1}(\bx)\sqrt{\mathbf A}_m(\bx)-\bm\Delta^\dag_m\mathbf B_m(\bx)\bm\Delta_m
\label{complete_square}
\end{align}
where $\bm\Delta_m = \mathbf Y_m-\mathbf B^{-1}_m(\bx)\sqrt{\mathbf A}_m(\bx)$.
We then obtain the optimal $\mathbf Y^\star_m=\mathbf
B^{-1}_m(\bx)\sqrt{\mathbf A}_m(\bx)$. Substituting this $\mathbf Y^\star_m$ in
$\tilde f_q$ recovers the original problem. 
\end{IEEEproof}

The quadratic transform for FP is first developed for the scalar case in
Proposition \ref{theorem:quadratic}, then generalized to the vector case
in \cite{part1}, where the objective function has the form
$\sum^M_{m=1}
f_m(\mathbf{a}^\dag_m(\bx)\mathbf{B}_m^{-1}(\bx)\mathbf{a}_m(\bx))$, where
$\mathbf a_m(\bx)\in\mathbb C^n$, $\mathbf B_m(\bx)\in\mathbb S^{n\times
n}_{++}$, and $f_m(z): \mathbb R\mapsto\mathbb R$ is a nondecreasing function.

The above result is a further generalization to the matrix case.
The scalar FP can be used to model the power control problem for
single-antenna links. The vector FP can be used to deal with a special case of
MIMO communication \cite{part1} where each link has at most one data stream.
To reap the full benefit of MIMO, each link needs to carry multiple data
streams. In this case, the matrix FP is necessary.


\begin{theorem}[Matrix Lagrangian Dual Transform]
\label{theorem:lagrangian_matrix}
Given a nonempty constraint set $\mathcal X$ as well as a sequence of numerator functions $\mathbf A_m(\bx)\in\mathbb H_+^{n\times n}$, denominator functions $\mathbf B_m(\bx)\in\mathbb H^{n\times n}_{++}$, and nonnegative weights $w_m\ge0$, for $m=1,2,\ldots,M$, the \emph{sum-of-weighted-logarithmic-matrix-ratios} problem
\begin{subequations}
\label{prob:lagrangian_matrix}
\begin{align}
\underset{\bx}{\text{maximize}} &\quad
\sum^M_{m=1}w_m\log\left|\bI+\sA^\dag(\bx)\mathbf{B}_m^{-1}(\bx)\sA(\bx)\right|
    \label{prob:lagrangian_matrix_obj}\\
\text{subject to}&\quad \mathbf{x}\in\mathcal X
\end{align}
\end{subequations}
is equivalent to
\begin{subequations}
\begin{align}
\label{}
\underset{\bx,\underline{\bm\Gamma}}{\text{maximize}}\quad&
f_r(\bx,\underline{\bm\Gamma})\\
\text{subject to}\quad& \bx\in\mathcal X,\\
&\bm\Gamma_m\in \mathbb H^{n\times n}_{+},
\end{align}
\end{subequations}
where the new objective function is
\begin{align}
\label{lagrangian_matrix:fr}
&f_r(\bx,\underline{\bm\Gamma})=\sum^M_{m=1}w_m\Big(\log\left|\bI+\bm\Gamma_m\right|-\text{tr}(\bm\Gamma_m)+\text{tr}\Big((\bI+\bm\Gamma_m)\notag\\
&\qquad\cdot\sA^\dag(\bx)\big(\mathbf A_m(\bx)+\mathbf B_m(\bx)\big)^{-1}\sA(\bx)\Big)\Big).
\end{align}
\end{theorem}
\begin{IEEEproof}
First, using the Woodbury matrix identity, i.e., $(\mathbf D+\mathbf U \mathbf C \mathbf V)^{-1}=\mathbf D^{-1}-\mathbf D^{-1}\mathbf U(\mathbf C^{-1}+\mathbf V\mathbf D^{-1}\mathbf U)^{-1}\mathbf V\mathbf D^{-1}$, we can rewrite (\ref{lagrangian_matrix:fr}) as
\begin{align}
\label{lagrangian_matrix:fr_orig}
&f_r(\bx,\underline{\bm\Gamma})=\sum^M_{m=1}w_m\Big(\log\left|\bI+\bm\Gamma_m\right|+n\,-\notag\\
&\;\,\text{tr}\Big((\bI+\bm\Gamma_m)\big(\bI+\sA^\dag(\bx)\mathbf B^{-1}_m(\bx)\sA(\bx)\big)^{-1}\Big)\Big).
\end{align}
We then consider the optimization of the above new form of $f_r$. Note that the optimization over $\bm\Gamma_m$ can be done separately for each term of the summation. Since each of the terms
is concave over $\bm\Gamma_m$ when $\bx$ is fixed, the optimal $\bm\Gamma_m$
can be determined by setting $\partial f_r/\partial \bm\Gamma_m$ to zero, i.e.,
\begin{equation}
(\bI + \bm\Gamma_m)^{-1} - \Big(\bI+\sA^\dag(\bx)\mathbf B^{-1}_m(\bx)\sA(\bx)\Big)^{-1} = \mathbf{0}.
\end{equation}
{\color{black}(Note that the derivative $\partial f_r/\partial \bm\Gamma_m$ exists in this case because $f_r$ is a spectral function \cite{lewis_spectral_func}.)}
Thus, we obtain the optimal $\bm\Gamma^\star_m=\sA^\dag(\bx)\mathbf B^{-1}_m(\bx)\sA(\bx)$.
Substituting this $\bm\Gamma^\star_m$ in (\ref{lagrangian_matrix:fr_orig}) recovers the original problem, thereby establishing the theorem.
\end{IEEEproof}

\setcounter{equation}{21}
\begin{figure*}[b]
\hrule
\begin{equation}
f_q(\bx,\underline{\bm\Gamma},\underline{\mathbf Y})=\sum^M_{m=1}\Big(w_m\log\left|\bI+\bm\Gamma_m\right|-w_m\text{tr}\big(\bm\Gamma_m\big)+\text{tr}\Big((\bI+\bm\Gamma_m)\Big(
2\sqrt{w_m}\sqrt{\mathbf A}_m^\dag(\bx)\mathbf Y_m-\mathbf Y^\dag_m\big(\mathbf A_m(\bx)+\mathbf{B}_m(\bx)\big)\mathbf Y_m\Big)\Big)\Big).
\label{joint:fq}
\end{equation}
\end{figure*}
\setcounter{equation}{20}

Observe that the proposed matrix quadratic transform of Theorem \ref{theorem:quadratic_matrix} can be applied to decouple the
ratio terms of $f_r$ in (\ref{lagrangian_matrix:fr}) to further transform
the matrix FP, as stated in the corollary below.
\begin{corollary}
\label{corollary:joint}
The problem (\ref{prob:lagrangian_matrix}) is equivalent to
\begin{subequations}
\begin{align}
\label{}
\underset{\bx,\underline{\bm\Gamma},\underline{\mathbf Y}}{\text{maximize}}\quad&
f_q(\bx,\underline{\bm\Gamma},\underline{\mathbf Y})\\
\text{subject to}\quad& \bx\in\mathcal X,\\
&\bm\Gamma_m\in \mathbb H^{n\times n}_{+},\\
&\mathbf Y_m\in\mathbb C^{n\times n},
\end{align}
\end{subequations}
where the new objective function is displayed in (\ref{joint:fq}) at the bottom of the next page. Note that $\Re\{\cdot\}$ can be dropped for the term $\sqrt{\mathbf A}_m^\dag(\bx)\mathbf Y_m$ because of trace.
\end{corollary}
\begin{IEEEproof}
Treating $f_m(\mathbf Z)=\text{tr}\big((\bI+\bm\Gamma_m)\mathbf Z\big)$ as the nondecreasing function, $\sqrt{w_m}\sA(\bx)$ as the square root of the numerator, and $\mathbf A_m(\bx)+\mathbf B_m(\bx)$ as the denominator, we apply the matrix quadratic transform of Theorem \ref{theorem:quadratic_matrix} to the last term of $f_r$ in (\ref{lagrangian_matrix:fr}) to obtain the above reformulation.
\end{IEEEproof}

Note that the new objective function $f_q$ is a linear function of each of the
square-root terms of the numerator $\sqrt{w_m}$ and $\sqrt{\mathbf A}_m(\bx)$ and the denominator term
$\mathbf B_m(\bx)$, while keeping all other terms fixed. This facilitates algorithm design for solving the matrix FP problem.
We also remark that there are also other ways of applying the matrix quadratic
transform to $f_r$ in (\ref{lagrangian_matrix:fr}) by choosing different matrix
ratios and functions $f_m(\cdot)$. The advantage of the above decomposition as
compared to the alternatives is discussed in details in Section \ref{subsec:pattern}.

{\color{black}
\section{Fractional Programming Transform as Minorization Maximization}
}
\label{sec:MM}

\begin{figure}[t]
\psfrag{D}[][]{\small$\hat{\bx}_{t+1}$}
\psfrag{E}[][]{\small$\hat{\bx}_t$}
\psfrag{F}[][]{\small$\hat{\bx}_{t-1}$}
\psfrag{B}[][]{\small$g(\mathbf{x}|\hat{\bx}_t)$}
\psfrag{C}[][l]{\small$g(\mathbf{x}|\hat{\bx}_{t-1})$}
\psfrag{A}[][]{\small$f(\mathbf{x})$}
\begin{minipage}[b]{1.0\linewidth}
\centering
\centerline{\includegraphics[width=8.0cm]{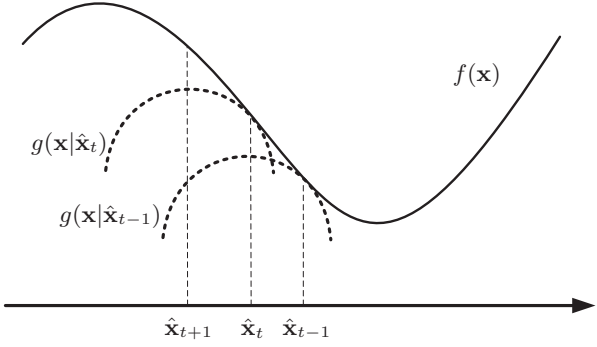}}
\caption{The iterative optimization by the MM algorithm. Observe that $f(\hat{\bx})$ is monotonically nondecreasing after each iteration.}
\label{fig:MM}
\end{minipage}
\end{figure}

An important theoretical observation made in this paper is that the matrix FP
transform proposed above can be recast in the MM framework.
First, we give a brief introduction to MM.
Consider a general optimization problem:
\setcounter{equation}{22}
\begin{subequations}
\label{prob:f}
\begin{align}
\underset{\bx}{\text{maximize}}\quad&
f(\bx)\\
\text{subject to}\quad& \bx\in\mathcal X
\end{align}
\end{subequations}
where $f(\bx)$ is not assumed to be concave.
Because of the nonconvexity, it is not always easy to solve the problem
directly. The core idea behind the MM algorithm is to successively solve a
sequence of \emph{well-chosen} approximations of the original problem
\cite{Razaviyayn,palomar}. Specifically, at point $\hat{\bx}\in\mathcal X$, the MM
algorithm approximates problem (\ref{prob:f}) as
\begin{subequations}
\label{prob:g}
\begin{align}
\underset{\bx}{\text{maximize}}\quad&
g(\bx|\hat{\bx})\\
\text{subject to}\quad& \bx\in\mathcal X
\end{align}
\end{subequations}
where $g(\bx|\hat{\bx})$ is referred to as the \emph{surrogate function} and is defined by the following two conditions:
\begin{itemize}
    \item{C1:} $g(\bx|\hat{\bx})\le f(\bx)$ for any $\bx\in\mathcal X$; 
    \item{C2:} $g(\hat{\bx}|\hat{\bx})=f(\hat{\bx})$. 
\end{itemize}
The MM algorithm updates $\hat{\bx}$ iteratively as follows:
\begin{equation}
\label{MM:maximization}
	\hat{\bx}_{t+1} = \arg\max_{\bx\in\mathcal X}g(\bx|\hat{\bx}_t)
\end{equation}
where subscript $t$ is the iteration index. Note that the function value of $f(\hat \bx)$ is nondecreasing after each iteration because
\begin{equation}
\label{converge:MM}
f(\hat{\bx}_{t+1})\overset{(a)}{\ge} g(\hat{\bx}_{t+1}|\hat{\bx}_t)\overset{(b)}{\ge} g(\hat{\bx}_t|\hat{\bx}_{t})\overset{(c)}{=}f(\hat{\bx}_t)
\end{equation}
where $(a)$ follows by C1, $(b)$ follows by the optimality of $\hat{\bx}_{t+1}$ in (\ref{MM:maximization}), and $(c)$ follows by C2. This is illustrated in Fig.~\ref{fig:MM} on the next page.

The following proposition gives a convergence analysis of the MM algorithm.
\begin{proposition}
\label{prop:stationary}
Let $\hat\bx_t$ be the solution produced by the MM update (\ref{MM:maximization}) after $t$ iterations. The function value $f(\hat\bx_t)$ has a non-decreasing convergence with $t$. Further, variable $\hat\bx_t$ converges to a stationary point solution to the original optimization problem (\ref{prob:f}) if the following three conditions are satisfied: (i) $f(\bx)$ is continuous over a convex closed set $\mathcal X$; (ii) $g(\bx|\hat\bx)$ is continuous in $(\bx,\hat\bx)$; (iii) $f(\bx)$ and $g(\bx|\hat\bx)$ are differentiable with respect to $\bx$ given $\hat\bx$.
\end{proposition}
\begin{IEEEproof}
The non-decreasing convergence of $f(\hat\bx)$ is already verified in (\ref{converge:MM}). Further, combining the above condition (iii) with the conditions C1 and C2, we obtain that $f(\bx)$ and $g(\bx|\hat\bx)$ have the same gradient with respect to $\bx$ at $\bx=\hat\bx$. This result, along with the above conditions (i) and (ii), guarantees that $\hat\bx_t$ converges to a stationary point solution to the original optimization problem (\ref{prob:f}) according to \cite{Razaviyayn}. We remark that the proof is adaptable to case where $\mathbf x$ is a complex variable; the argument is similar to that of \cite{song_babu_palomar}.
\end{IEEEproof}

The MM algorithm is a framework rather than an algorithmic prescription,
because the algorithm depends on the specific choice of the surrogate function.
If $f(\cdot)$ is twice differentiable, its second order Taylor expansion is often the
first candidate to check to see whether it is suitable as a surrogate function.
For more general functions, many of the ingenious
ways of constructing a surrogate function have been documented in \cite{palomar}.

The main point of this section is that the proposed matrix FP transforms can
actually be interpreted in the MM framework as constructing surrogate functions
of the original problems, as stated below.

\begin{figure*}[hb]
\setcounter{equation}{31}
\hrulefill
\begin{subequations}
\begin{align}
&f_q(\underline{s},\underline{\mathbf{V}},\underline{\bm\Gamma},\underline{\mathbf Y})\notag\\
    &=\sum_{j\in\mathcal J}\Big(w_{j s_j}\log\left|\bI+\bm\Gamma_j\right|-w_{j s_j}\text{tr}\big(\bm\Gamma_j\big)+\text{tr}\Big((\bI+\bm\Gamma_j)\Big(2\sqrt{w_{js_j}}\,\mathbf{H}_{js_j}\mathbf{V}_{s_j}\mathbf Y^\dag_j-
\mathbf Y^\dag_j\big(\mathbf F_{j}+\mathbf{H}_{js_j}\mathbf{V}_{s_j}\mathbf{V}^\dag_{s_j}\mathbf{H}^\dag_{js_j}\big)\mathbf Y_j\Big)\Big)\Big)
        \label{fq}\\
    &=\sum_{j\in\mathcal J}\bigg(w_{j s_j}\log\left|\bI+\bm\Gamma_j\right|-w_{j s_j}\text{tr}\big(\bm\Gamma_j\big)+\text{tr}\Big( 2\sqrt{w_{js_j}}\,(\bI+\bm\Gamma_j)\mathbf{H}_{js_j}\mathbf{V}_{s_j}\mathbf Y^\dag_j-\sum_{j'\in\mathcal J}(\bI+\bm\Gamma_{j'})\mathbf Y^\dag_{j'}\mathbf{H}_{j's_j}\mathbf{V}_{s_j}\mathbf{V}^\dag_{s_j}\mathbf{H}^\dag_{j's_j}\mathbf Y_{j'}\Big)\bigg)\notag\\
    &\qquad+\sum_{j\in\mathcal J} \sigma^2\mathbf Y^\dag_j(\bI+\bm\Gamma_j)\mathbf Y_j.
        \label{fq_new_form}
\end{align}
\end{subequations}
\end{figure*}
\setcounter{equation}{26}

\begin{theorem}
\label{remark:quadratic_MM}
Consider the matrix quadratic transform in Theorem \ref{theorem:quadratic_matrix},
if we consider the optimal ${\mathbf Y}_m^\star$ as a function of $\hat\bx$ and substitute it into $\tilde f_q$, then the new objective function $\tilde f_q(\bx,\underline{\mathbf Y}(\hat\bx))$, where
\begin{equation}
\mathbf Y_m(\hat{\bx})=\mathbf B^{-1}_m(\hat\bx)\sA(\hat\bx),
\end{equation}
is a surrogate
function of the objective of the optimization problem (\ref{prob:quadratic_matrix}).
\end{theorem}

\begin{IEEEproof}
We use $f_\text{I}(\bx)$ to denote the objective function in
(\ref{prob:quadratic_matrix_fo}). Substitute $\mathbf Y_m(\hat\bx)=\mathbf
B^{-1}_m(\hat\bx)\sA(\hat\bx)$ back in $\tilde{f}_q$.
We aim to show that $g(\bx|\hat\bx)=\tilde{f}_q(\bx, \underline{\mathbf
Y}(\hat{\mathbf x}))$ is a surrogate function of $f_\text{I}(\bx)$.

As already shown in the proof of Theorem \ref{theorem:quadratic_matrix},
$\underline{\mathbf{Y}}({\mathbf x})$ is the optimum solution for the
maximization of $\tilde f_q(\bx,\underline {\mathbf Y})$ over
$\underline{\mathbf Y}$ when $\mathbf x$ is fixed. So,
$\tilde{f}_q(\bx, \underline{\mathbf Y}(\hat{\mathbf x}))\le\tilde{f}_q(\bx,
\underline{\mathbf Y}({\mathbf x})), \forall \hat \bx, \bx$.
Further, it can be seen that
$\tilde{f}_q(\bx, \underline{\mathbf Y}({\mathbf x}))=f_\text{I}({\mathbf x})$
for any $\bx$.

Thus, for each fixed $\hat \bx$, we have $\tilde{f}_q(\bx, \underline{\mathbf
Y}(\hat{\mathbf x}))\le f_\text{I}(\bx)$, $\forall \bx$, and $\tilde{f}_q(\hat\bx,
\underline{\mathbf Y}(\hat{\mathbf x}))=f_\text{I}(\hat\bx)$, thus verifying
the conditions C1 and C2 for $\tilde{f}_q(\bx,\underline{\mathbf Y}(\hat\bx))$ to
be a surrogate function of $f_\text{I}(\bx)$.
\end{IEEEproof}

\begin{theorem}
\label{remark:lagrangian_MM}
Consider the matrix Lagrangian dual transform in Theorem \ref{theorem:lagrangian_matrix},
if we consider the optimal ${\bm\Gamma}_m^\star$ as a function of $\hat\bx$ and substitute it into $f_r$, then the new objective function $f_r(\bx,\underline{\bm\Gamma}(\hat\bx))$, where
\begin{equation}
\bm\Gamma_m(\hat{\bx})=\sA^\dag(\hat{\bx})\mathbf B^{-1}_m(\hat{\bx})\sA(\hat{\bx}),
\end{equation}
is a surrogate
function of the objective of the optimization problem (\ref{prob:lagrangian_matrix}).
\end{theorem}

\begin{IEEEproof}
We use $f_\text{II}(\bx)$ to denote the objective function in
(\ref{prob:lagrangian_matrix_obj}). We substitute
$\bm\Gamma_m(\hat\bx)=\sA^\dag(\hat{\bx})\mathbf B^{-1}_m(\hat{\bx})\sA(\hat{\bx})$ back in $f_r$,
and aim to show that
$g(\bx|\hat\bx)=f_r(\bx,\underline{\bm\Gamma}(\hat{\bx}))$ is a surrogate
function of $f_\text{II}(\bx)$.

As shown in the proof of Theorem \ref{theorem:lagrangian_matrix},
$\underline{\bm\Gamma}(\bx)$ is the optimal solution
to maximizing $f_r(\bx, \underline{\mathbf \Gamma})$ over $\underline{\mathbf
\Gamma}$ when $\bx$ is fixed, so
$f_r(\bx,\underline{\bm\Gamma}(\hat\bx))\le
f_r(\bx,\underline{\bm\Gamma}(\bx)), \forall \bx, \hat\bx$. Also, it holds true that
$f_r(\hat{\bx},\underline{\bm\Gamma}(\hat{\bx}))=f_\text{II}(\hat{\bx})$, $\forall \hat{\bx}$. Combining
the above results, we see that the
conditions C1 and C2 are satisfied, thus $f_r(\bx,\underline{\bm\Gamma}(\hat{\bx}))$ is a surrogate
function of $f_\text{II}(\bx)$.
\end{IEEEproof}

\begin{corollary}
\label{remark:joint_MM}
Consider the transform in Corollary \ref{corollary:joint},
if we consider the optimal ${\bm\Gamma}_m^\star$ and the optimal $\mathbf Y^\star_m$ as two functions of $\hat\bx$, and substitute them into into $f_q$,
then the new objective function $f_q(\bx,\underline{\bm\Gamma}(\hat\bx),\underline{\mathbf Y}(\hat\bx))$,
where
\begin{equation}
\bm\Gamma_m(\hat\bx)=\sA^\dag(\hat\bx)
\mathbf B_m^{-1}(\hat\bx)
\sA(\hat\bx)
\end{equation}
and
\begin{equation}
\mathbf Y_m(\hat\bx)=
\big(\mathbf A_m(\hat\bx)+\mathbf B_m(\hat\bx)\big)^{-1}
\big(\sqrt{w_m}\sqrt{\mathbf A}_m(\hat\bx)\big),
\end{equation}
is a surrogate
function of the objective of the optimization problem (\ref{prob:lagrangian_matrix}).
\end{corollary}
\begin{IEEEproof}
Again, let $f_\text{II}(\bx)$ be the objective function in
(\ref{prob:lagrangian_matrix_obj}). We introduce two new variables
$\hat\bx$ and $\hat{\hat\bx}$, and substitute
$\bm\Gamma_m(\hat\bx)=\sA^\dag(\hat\bx) \mathbf B^{-1}_m(\hat{{\bx}}) \sA(\hat\bx)$ and
$\mathbf Y_m(\hat{\hat\bx})=
\big(\mathbf A_m(\hat{\hat \bx})+\mathbf{B}_m(\hat{\hat\bx})\big)^{-1}
\big(\sqrt{w_m}\sA(\hat{\hat{\bx}})\big)$ back in ${f}_q$ and $f_r$.
Let
$g_1(\bx|\hat{\hat\bx},\hat\bx)=f_q(\bx,\underline{\bm\Gamma}(\hat\bx),\underline{\mathbf
Y}(\hat{\hat\bx}))$, and
$g_2(\bx|\hat\bx) = f_r(\bx,\underline{\mathbf \Gamma}(\hat\bx))$.


According to Theorem \ref{remark:lagrangian_MM}, $g_2(\bx|\hat\bx)$ is
a surrogate function of $f_\text{II}(\bx)$ in the sense that
$g_2(\bx|\hat\bx)\le f_\text{II}(\bx)$ and $g_2(\hat\bx|\hat\bx)=
f_\text{II}(\hat\bx), \forall \bx, {\hat\bx}$.
According to Theorem
\ref{remark:quadratic_MM}, $g_1(\bx|\hat{\hat{\bx}},\hat{\bx})$ is a surrogate function with respect to $f_r$
in the sense that $g_1(\bx|\hat{\hat\bx},\hat \bx)\le f_r(\bx,\underline{\mathbf \Gamma}(\hat\bx))$ and
$g_1(\hat{\hat\bx}|\hat{\hat\bx},\hat\bx)= f_r(\hat{\hat\bx},\underline{\mathbf \Gamma}(\hat{\bx})),
\forall \bx, \hat\bx, \hat{\hat\bx}$.

Combining these results and fixing $\hat{\hat\bx}= \hat\bx$,
we obtain $g_1(\bx|\hat \bx,\hat \bx) \le f_r(\bx, \underline{\mathbf \Gamma}(\hat\bx))
= g_2 (\bx|\hat\bx) \le f_\text{II}(\bx), \forall \bx$ and
$g_1(\hat\bx|\hat \bx,\hat \bx) = f_r(\hat\bx, \underline{\mathbf \Gamma}(\hat \bx))=
g_2(\hat\bx|\hat\bx) = f_\text{II}(\hat\bx)$, thereby verifying
the conditions C1 and C2 for
$f_q(\bx,\underline{\bm\Gamma}(\hat{\bx},\underline{\mathbf Y}({\hat\bx}))$ to
be a surrogate function of $f_\text{II}(\bx)$.
\end{IEEEproof}

This MM interpretation of the FP transforms provides a theoretical basis
for the proposed FPLinQ strategy for joint scheduling, beamforming and power
control.
Note that the above results carry over to Propositions \ref{theorem:quadratic} and
\ref{theorem:lagrangian} for the scalar FP case, so the approach of \cite{part1,part2} can be interpreted as an MM algorithm as well. 

\section{Joint Scheduling, Power Control and Beamforming Using FPLinQ}
\label{sec:approach}

\begin{figure*}[hb]
\hrulefill
\setcounter{equation}{37}
\begin{equation}
\label{lambda}
\lambda_{ji}=w_{ji}\log\left|\bI+\bm\Gamma_j\right|-w_{ji}\text{tr}(\bm\Gamma_j)
    + \text{tr}
    \bigg(2\sqrt{w_{ji}}\,(\bI+\bm\Gamma_j)\mathbf Y^\dag_j\mathbf{H}_{ji}\tilde{\mathbf{V}}_{ji}-
    \sum_{j'\in\mathcal J}(\bI+\bm\Gamma_{j'})\mathbf Y^\dag_{j'}\mathbf{H}_{j'i}\tilde{\mathbf{V}}_{ji}\tilde{\mathbf{V}}^\dag_{ji}\mathbf{H}^\dag_{j'i}\mathbf Y_{j'}\bigg).
\end{equation}
\end{figure*}

\subsection{Iterative Optimization via Matrix FP}
\label{subsec:FPLinQ}


We propose to solve the joint scheduling and beamforming problem (\ref{prob:d2d})
iteratively by first reformulating it using Corollary \ref{corollary:joint}.
Specifically, after specializing the variable $\bx$ in
(\ref{prob:lagrangian_matrix}) to be the $(\underline{\mathbf V},\underline s)$
in (\ref{prob:d2d}), we obtain the following reformulation:
\begin{theorem}
\setcounter{equation}{30}
\label{prop:d2d_transform}
The joint beamforming and link scheduling problem (\ref{prob:d2d}) is equivalent to
\begin{subequations}
\label{prob:new}
\begin{align}
\underset{\underline{s},\underline{\mathbf V},\underline{\bm\Gamma},\underline{\mathbf Y}}{\text{maximize}}\quad&
f_q(\underline{s},\underline{\mathbf V},\underline{\bm\Gamma},\underline{\mathbf Y})\\
\text{subject to}\quad& \text{(\ref{prob:V}), (\ref{prob:s_in}), (\ref{prob:s_ne})},\notag\\
&\bm\Gamma_j\in \mathbb H^{N\times N}_{+},\\
&\mathbf Y_j\in\mathbb C^{N\times N},
\end{align}
\end{subequations}
where the new objective function $f_q$ is shown in (\ref{fq}) as displayed at the bottom of the page.
\end{theorem}

\begin{IEEEproof}
The reformulating steps directly follow Corollary \ref{corollary:joint}. We remark that $f_q$ can be rewritten as in (\ref{fq_new_form}), which enables an efficient optimization by matching.
\end{IEEEproof}

We now optimize over the variables of the new problem (\ref{prob:new}) in an iterative manner. First, when $\underline s$ and $\underline{\mathbf V}$ are both held fixed, the auxiliary variables $\underline{\bm\Gamma}$ and $\underline{\mathbf{Y}}$ can be optimally determined as
\setcounter{equation}{32}
\begin{equation}
\label{opt_Gamma}
    \bm\Gamma^\star_j =
\mathbf{V}^\dag_{s_j}\mathbf{H}^\dag_{js_j}\mathbf F^{-1}_{j}\mathbf{H}_{js_j}\mathbf{V}_{s_j}
\end{equation}
and
\begin{equation}
\label{opt_Y}
\mathbf Y^\star_j =\left(\mathbf F_{j}+\mathbf{H}_{js_j}\mathbf{V}_{s_j}\mathbf{V}^\dag_{s_j}\mathbf{H}^\dag_{js_j}\right)^{-1}\sqrt{w_{js_j}}\,\mathbf{H}_{js_j}\mathbf{V}_{s_j}.
\end{equation}
We remark that the implicit constraints as stated in Theorem \ref{theorem:quadratic_matrix} are automatically satisfied by the above optimal solution of the auxiliary variable $\mathbf Y^\star_j$.

It remains to optimize the beamforming variable $\underline{\mathbf V}$ and
the scheduling variable $\underline{s}$. The key idea is to formulate the
problem as a bipartite weighted matching problem, which is described in detail
below.

{\color{black}
\subsection{Scheduling and Beamforming via Bipartite Matching}
}

We consider the objective function $f_q$ of the form (\ref{fq_new_form}). The
key observation is that the beamformer of each link (if it is scheduled) can be
optimally determined from $f_q$, even without knowing the
scheduling decisions for the nearby links. To formalize this idea, let
$\tilde{\mathbf V}_{ji}$ be the tentative value of $\mathbf V^\star_i$ if
link $(i,j)$ is scheduled. By completing the square in $f_q$, we can compute
$\tilde{\mathbf V}_{ji}$ as
\begin{multline}
\label{tilde_V}
\tilde{\mathbf{V}}_{ji} = \bigg(\mathbf \mu_{ji}\bI+\sum_{j'\in\mathcal J}\mathbf{H}^\dag_{j'i}\mathbf{Y}_{j'}(\bI+\bm\Gamma_{j'})\mathbf{Y}^\dag_{j'}\mathbf{H}_{j'i}\bigg)^{-1}\\
\cdot\sqrt{w_{ji}}\,\mathbf{H}^\dag_{ji}\mathbf{Y}_{j}(\bI+\bm\Gamma_j)
\end{multline}
where $\mu_{ji}$ is a Lagrangian multiplier for the power constraint (\ref{prob:V}), optimally determined as
\begin{equation}
\mu_{ji}^\star = \min\{\mu_{ji}\ge0: \text{tr}( \tilde{\mathbf{V}}_{ji}^\dag\tilde{\mathbf{V}}_{ji})\le P_{\max}\},
\end{equation}
which can be computed efficiently by bisection search since $\tilde{\mathbf V}_{ji}$ is monotonically decreasing with $\mu_{ji}$. The solution $\tilde{\mathbf V}_{ji}$ in (\ref{tilde_V}) has the same structure as an MMSE beamformer.

We now turn to the question of which $\tilde{\mathbf V}_{ji}$ should be
chosen to be $\mathbf V_i$ so as to maximize $f_q$. This is akin to a
scheduling step of choosing the best transmitter $i$ for each receiver $j$.
The key is to recognize this question as a weighted bipartite
matching problem:
\setcounter{equation}{36}
\begin{subequations}
\label{prob:matching}
\begin{align}
\underset{\underline q}{\text{maximize}}\quad&
\sum_{j\in\mathcal J}\sum_{i\in\mathcal K_j}\lambda_{ji}q_{ji}\\
\text{subject to}\quad& \sum_{i\in\mathcal K_j}q_{ji}\le1,\\
& \sum_{j\in\mathcal L_i}q_{ji} \le 1,\\
& q_{ji}\in\{0,1\},\\
& q_{ji}=0 \;\;\text{if}\; i\notin\mathcal K_j \; \text{or} \; j\notin\mathcal L_i,
\end{align}
\end{subequations}
where the weight $\lambda_{ji}$ is evaluated by (\ref{lambda}) as displayed at
the bottom of the page, and $q_{ji}$ is the matching variable between the associated transmitters and receivers. This weighted bipartite matching problem can be solved
optimally in polynomial time by using well-known approaches such as the
Hungarian \cite{hungarian} or the auction \cite{auction} algorithm.

Note that (\ref{prob:matching}) is typically a \emph{sparse} matching problem,
since most pairs of $(i,j)\in\mathcal I\times\mathcal J$ are not associated, so
the auction algorithm is likely to be more efficient than the Hungarian
algorithm. The matching variable $q_{ji}$ indicates $\mathbf V_i$ should be
set to which of the $\tilde{\mathbf V}_{ji}$.
Mathematically, $\underline{\mathbf V}$ is recovered as
\setcounter{equation}{38}
\begin{equation}
\label{opt_V}
	\mathbf V^\star_i =
\left\{
\begin{aligned}
&\tilde{\mathbf V}_{ji},\;\text{if }\; q_{ji}=1 \; \text{for some } j;\\
&\mathbf{0},\quad\; \text{otherwise}.
\end{aligned}
\right.
\end{equation}

After updating $\underline{\mathbf V}$, the final step is to update the
scheduling variable $\underline s$ for the fixed $\underline{\mathbf V}$.
This is again a weighted bipartite matching problem, but now since
${\mathbf V}_i$ is fixed, this amounts to choosing the best receiver $j$
for each transmitter $i$:
\begin{subequations}
\label{prob:matching2}
\begin{align}
\underset{\underline q}{\text{maximize}}\quad&
\sum_{i\in\mathcal I}\sum_{j\in\mathcal L_i} q_{ji} w_{ji} r_{ji}\\
\text{subject to}\quad& \sum_{i\in\mathcal K_j} q_{ji}\le1,\\
& \sum_{j\in\mathcal L_i}q_{ji} \le 1,\\
& q_{ji}\in\{0,1\},\\
& q_{ji}=0 \;\;\text{if}\; i\notin\mathcal K_j \; \text{or} \; j\notin\mathcal L_i,
\end{align}
\end{subequations}
where $w_{ji} r_{ji}$ is the weighted achievable rate if the receiver $j$ is
scheduled for transmitter $i$ under fixed ${\mathbf V}_i$.
Note that since all
${\underline {\mathbf V}}$ are fixed, $r_{ij}$ can be computed independently of
the schedule, using an expression similar to (\ref{rate}).
This problem can again be solved in polynomial time.
The optimal schedule is then determined from the optimal
$q_{ij}$ as
\begin{equation}
\label{opt_s}
	s^\star_j =
\left\{
\begin{aligned}
& i ,\quad\;\,\text{if }\; q_{ji}=1 \; \text{for some } i;\\
& \varnothing , \quad \text{otherwise}.
\end{aligned}
\right.
\end{equation}
We note that the reason for having two sets of matching is because we allow a
general network model in which each transmitter may associate with multiple
receivers and each receiver may associate with multiple transmitters.
For simpler D2D model such as the one in Fig.~\ref{fig:topology}(a), these
two matching steps would not have been necessary, as in \cite{shen_isit}.

\begin{algorithm}[t]
Initialize all the variables to feasible values\;
\Repeat{the weighted sum rate converges}{
    Update $\underline{\bm\Gamma}$ according to (\ref{opt_Gamma})\;
    Update $\underline{\mathbf{Y}}$ according to (\ref{opt_Y})\;
    Update $\underline{\mathbf{V}}$ according to (\ref{opt_V})\;
    Update $\underline{s}$ by weighted bipartite matching (\ref{opt_s});
}
\caption{Proposed FPLinQ Strategy for D2D Link Scheduling with Power Control and Beamforming}
\label{alg:FPLinQ}
\end{algorithm}

\begin{table*}[t]
\renewcommand{\arraystretch}{1.3}
\small
\centering
{\color{black}
\caption{\small Comparison of Link Scheduling Algorithms for D2D Networks}
\begin{tabular}{|c||c|c|c|c|c|}
\hline
& FPLinQ &  FlashLinQ \cite{FlashLinQ} & ITLinQ \cite{ITLinQ} & ITLinQ+
\cite{ITLinQ+} & BCD\cite{weiyu_TWC} \\
\hline
\hline
Scheduling \& Association & Flexible & Single & Single & Single & Flexible \\
\hline
Power Control & $\checkmark$ & \textsf{x} & \textsf{x} & $\checkmark$ & $\checkmark$\\
\hline
Beamforming & $\checkmark$ & \textsf{x} & \textsf{x} & \textsf{x} & $\checkmark$ \\
\hline
Tuning Parameters & Not Needed & Required & Required & Required & Not Needed \\
\hline
Convergence with Fixed Schedule & Stationary Point &  -- &  -- &  -- & Stationary Point\\
\hline
Computational Complexity & $O(L^2(N^4+\log L))$ & $O(L^2)$ & $O(L^2)$ & $O(L^2)$ & $O(L^2(N^4+\log L))$\\
\hline
Communication Complexity & $O(N^2L^2)$ & $O(L^2)$ & $O(L^2)$ & $O(L^2)$ & $O(N^2L^2)$\\
\hline
Link Reactivation & $\checkmark$ & \textsf{x} & \textsf{x} & \textsf{x} & \textsf{x}\\
\hline
\end{tabular}
\label{tab:comparison}
}
\end{table*}

Combining all the above steps together yields the FPLinQ strategy.
Algorithm \ref{alg:FPLinQ} at the top of the page summarizes the overall approach.

A desirable feature of FPLinQ as compared to FlashLinQ, ITLinQ and ITLinQ+ is that no tuning of design parameters is needed. But, FPLinQ is also somewhat more difficult to implement in a distributed fashion than FlashLinQ, ITLinQ, and ITLinQ+, because it additionally requires the update of the auxiliary variables $\underline{\bm\Gamma}$ and $\underline{\mathbf Y}$ per iteration. \\

{\color{black} \subsection{Alleviating Premature Turn-Off}}

It is worthwhile to take a deeper look into Algorithm \ref{alg:FPLinQ} to
understand how FPLinQ is able to alleviate the \emph{premature turning-off}
problem. FPLinQ differs from the BCD method mainly in Step 5, where the
beamforming variable $\underline{\mathbf V}$ is optimized for the new objective
function $f_q$ instead of the weighted sum-rate objective. Taking a close look
at (\ref{prob:matching}) and (\ref{opt_V}), we can see that the update of each
$\mathbf V_i$ at Step 5 of FPLinQ is not affected by the current value of
$\underline s$. From the MM interpretation, we see that when updating
$\underline{\mathbf V}$ for fixed $\underline s$, FPLinQ is actually using
the surrogate function to mimic the original objective function so that the
optimization over $\underline{\mathbf V}$ no longer relies on $\underline s$.
In comparison to the BCD method, this less aggressive update of
$\underline{\mathbf V}$ by FPLinQ allows the existing \textsc{off}-transmitters
to be reactivated, thereby alleviating the premature turning-off issue as
mentioned in Section \ref{subsec:BCD}. \\

{\color{black}
\subsection{Convergence Analysis}}

We now examine the convergence behavior of the proposed algorithm
by utilizing the MM interpretation as a tool.

\setcounter{equation}{41}
\begin{figure*}[hb]
\hrulefill
\begin{equation}
    \label{fr}
    f_r(\underline{s},\underline{\mathbf{V}},\underline{\bm\Gamma}) = \sum_{j\in\mathcal J}w_{js_j}\left(\log\left|\bI+\bm\Gamma_j\right|-\text{tr}(\bm\Gamma_j)
    + \text{tr}\bigg((\bI+\bm\Gamma_j)\mathbf{V}^\dag_{s_j}\mathbf{H}^\dag_{js_j}\left(\mathbf F_{j}+\mathbf{H}_{js_j}\mathbf{V}_{s_j}\mathbf{V}^\dag_{s_j}\mathbf{H}^\dag_{js_j}\right)^{-1}\mathbf{H}_{js_j}\mathbf{V}_{s_j}\bigg)\right).
\end{equation}
\end{figure*}

\begin{theorem}
The weighted sum rate across all the D2D links is nondecreasing after each
iteration of Algorithm \ref{alg:FPLinQ}, so the objective function of the optimization problem is guaranteed to converge. Furthermore, the optimization variables also converge. Finally, at convergence, for fixed $\underline s$, the solution
$\underline{\mathbf V}$ is a stationary point of the problem (\ref{prob:d2d}).
\end{theorem}

\begin{IEEEproof}
We prove convergence based on the MM interpretation of the FP transforms.
The Step 3 and Step 4 of the algorithm construct the surrogate functions as
defined in Theorem \ref{remark:lagrangian_MM} and Theorem \ref{remark:quadratic_MM}.  Step 5 of Algorithm
\ref{alg:FPLinQ} performs the maximization step of the MM algorithm, so the
weighted sum rate must be nondecreasing after Step 5. Step 6 further optimizes
the link schedule, so the weighted sum rate is nondecreasing after Step 6.
Since the optimization objective
is nondecreasing and is bounded above, Algorithm \ref{alg:FPLinQ} must converge in objective value.

The weighted sum rate is a differentiable function over
$\underline{\mathbf V}$ under fixed $\underline{s}$, so at convergence the solution of
$\underline{\mathbf V}$ given by Algorithm \ref{alg:FPLinQ} must be a
stationary point according to Proposition \ref{prop:stationary}.
\end{IEEEproof}

We remark that proving the convergence of Algorithm \ref{alg:FPLinQ} without the MM interpretation would have been much more cumbersome.

\subsection{Complexity Analysis}
\label{subsec:complexity}

We now analyze the complexity of FPLinQ (i.e., Algorithm \ref{alg:FPLinQ}).
We assume that there are a total of $L$ D2D links in the network; each
transmitter/receiver is associated with a small number (i.e., constant number)
of neighboring devices, so that $|\mathcal I| = O(L)$ and $|\mathcal J|=O(L)$.
To ease the analysis, we assume that FPLinQ runs for a fixed number of
iterations.

\emph{Communication Complexity:}  In each iteration of FPLinQ, each transmitter
$i$ requires the tuple $(\underline{\bm\Gamma},\underline{\mathbf Y},
\underline{s})$ to update $\mathbf V_i$, while every receiver $j$ requires
$\underline{\mathbf V}$ to update $\bm\Gamma_j$ and $\mathbf Y_j$. Each of
$\mathbf{V}_i, {\bm\Gamma}_j, {\mathbf Y}_j$ is an $N \times N$ matrix. Further,
the channel coefficients from $O(L^2)$ direct and interfering channels are
needed, with each channel being an $N \times N$ matrix. Thus, the total
communication complexity of these updates is $O(N^2L^2)$.  The two matchings in
Step 5 and Step 6 require the matching weights of all the links, thus
introducing a communication complexity of $O(L)$. 
The overall communication complexity of FPLinQ is then $O(N^2L^2)$. 
In the single-antenna single-association case, the communication complexity
of FPLinQ in each iteration is $O(L^2)$; in comparison, the communication
complexity of each step of FlashLinQ, ITLinQ, and ITLinQ+ is also $O(L^2)$,
as they all require the $O(L^2)$ channel coefficients.

\emph{Computational Complexity:} We first consider the update steps of FPLinQ
prior to matching, which as analyzed in \cite{part2} has a per-iteration
computational complexity of $O(N^4 L^2)$. The matching step can be performed
using the auction algorithm \cite{auction}, which has a computational complexity of
$O(L|\mathcal I|\log|\mathcal I|+L|\mathcal J|\log|\mathcal J|) = O(L^2 \log(L))$.
Thus, the overall per-iteration computational complexity of FPLinQ is $O(N^4L^2 + L^2\log(L))$. In the single-antenna single-association case, the
per-iteration computational complexity of FPLinQ reduces to $O(L^2\log L)$, while the
total computational complexities of FlashLinQ, ITLinQ, and ITLinQ+ are
all equal to $O(L^2)$.

We observe that the computational complexity of FPLinQ is sensitive to the
number of antennas $N$ (mainly due to the matrix inverse). Overall,
asymptotically, FPLinQ has the same communication complexity, but
higher computational complexity than the greedy based approaches---FlashLinQ, ITLinQ, and ITLinQ+. Note that although the joint scheduling and power control problem is NP-hard in general \cite{Sharma_Mazumdar_Shroff_MobiCom06,luo_zhang}, recent results nevertheless show that scalable implementation is feasible for a metropolitan-scale network with thousands of terminals \cite{Zhou_Guo_Mobihoc17,Zhou_Guo_arXiv}. In particular, \cite{Zhou_Guo_arXiv} uses the scalar FP method of \cite{part1,part2}.

{\color{black}
Table \ref{tab:comparison} summarizes the comparison between the proposed FPLinQ algorithm and the main benchmarks. The main advantage of FPLinQ is that it allows for flexible association, guarantees convergence without needing tuning parameters, while alleviating the potential pre-mature turn-off problem.
}

{\color{black}
\subsection{Different Ways to Decouple the Ratios}}
\label{subsec:pattern}

In the derivation of FPLinQ, we decouple the matrix ratios of $f_r$ shown in (\ref{fr}) at the bottom of the page in a
particular form, but such decoupling is not unique. There exist other ways to
decouple the ratio.

Recall that the proposed reformulation in Theorem \ref{prop:d2d_transform}
follows the proof of Corollary \ref{corollary:joint} by treating $(\bI+\bm\Gamma_m)$
as the fixed weight, with $f_m(\mathbf Z)=\text{tr}((\bI+\bm\Gamma_m)\mathbf Z)$,
where $\mathbf Z$ is a matrix ratio, i.e., $(\bI+\bm\Gamma) {\mathbf Z}$ is
\setcounter{equation}{42}
\begin{equation}
\label{pattern:proposed}
(\bI+\bm\Gamma)\,
\boxed{\sqrt{w}\mathbf{V}^\dag\mathbf{H}^\dag}\uwave{\left(\mathbf F+\mathbf{H}\mathbf{V}\mathbf{V}^\dag\mathbf{H}^\dag\right)}^{-1}\boxed{\sqrt{w}\mathbf{H}\mathbf{V}}
\end{equation}
as in (\ref{fr}). Here, the boxed component represents the numerator and the underlined component the denominator; all the subscripts are omitted for notational simplicity.

The matrix ratio in (\ref{pattern:proposed}) can also be decoupled in other ways. For instance, we could have included the term $(\bI+\bm\Gamma)$ in the numerator, i.e.,
\begin{equation}
\label{pattern:TSP}
\boxed{\sqrt{\mathbf A}^\dag}\uwave{\left(\mathbf F+\mathbf{H}\mathbf{V}\mathbf{V}^\dag\mathbf{H}^\dag\right)}^{-1}\boxed{\sqrt{\mathbf A}}
\end{equation}
where
\begin{equation}
\mathbf A = w\mathbf{H}\mathbf{V}(\bI+\bm\Gamma)\mathbf{V}^\dag\mathbf{H}^\dag.
\end{equation}
In fact, the above decoupling is exactly what \cite{part1} and \cite{part2} use
when treating scalar FP problems. However, the inclusion of the $(\bI + \bm\Gamma)$ term would result in an extra matrix decomposition step when computing the matrix square root, hence the resulting algorithm would be somewhat computationally more complex.

Alternatively, we could have excluded $w$ from $\mathbf A$, i.e.,
\begin{equation}
\label{pattern:WMMSE}
(\bI+\bm\Gamma)w\,\boxed{\mathbf{V}^\dag\mathbf{H}^\dag}\uwave{\left(\mathbf F+\mathbf{H}\mathbf{V}\mathbf{V}^\dag\mathbf{H}^\dag\right)}^{-1}\boxed{\mathbf{H}\mathbf{V}}\,.
\end{equation}
The above pattern yields yet another different $f_q$. It turns out that optimizing
$\underline{\mathbf V}$, $\underline{\bm\Gamma}$, and $\underline{\mathbf Y}$
iteratively for this particular $f_q$ recovers the WMMSE algorithm \cite{luo}
for beamforming. This connection to WMMSE has been shown for the vector FP case
in \cite{part2}. (As a corollary, this implies that there is a relation between
the WMMSE algorithm and the MM algorithm as well!) However, since $w_{js_j}$
contains the scheduling decision, this approach leads us to the situation that
$\mathbf V_i$'s are updated only for the \textsc{on}-links, thus it suffers
from the premature turning-off problem. \\

\section{Simulation Results}
\label{sec:simulation}

We validate the performance of FPLinQ through comparison with the benchmark
methods for a D2D network in a $1$km$\times$1km square area where the D2D links are randomly
located. Following \cite{FlashLinQ,ITLinQ,ITLinQ+}, we adopt the short-range
outdoor channel model ITU-1411 and use a 5MHz-wide frequency band centered at
$2.4$GHz. Moreover, the antenna height of each device is $1.5$m; the antenna gain
is $2.5$dBi; the noise power spectrum density is $-169$dBm/Hz; the noise figure is
7dB; the maximum transmit power is $20$dBm; the shadowing is modeled as a Gaussian
random variable in decibel with the standard deviation of $10$; the distance between the
transmitter and receiver of each link is uniformly distributed between $2$m and
$65$m.

The first simulation setting follows \cite{FlashLinQ,ITLinQ,ITLinQ+}: Given a
set of links with single-antenna transmitters/receivers and fixed single
association (as shown in Fig.~\ref{fig:topology}), the aim is to maximize the
sum rate across the links. We use FlashLinQ \cite{FlashLinQ}, ITLinQ
\cite{ITLinQ}, and ITLinQ \cite{ITLinQ+} as benchmarks. The BCD method
is equivalent to FPLinQ in this single-association case.
Because the benchmark methods do not have power control, for fair
comparison, we modify FPLinQ slightly to restrict the power to be
either zero or the maximum, i.e., round each $\mathbf V_i$ to $\{0,\sqrt{P_{\max}}\}$. This new version of FPLinQ
without power control is referred to as ``FPLinQ (no pc)''. Further, we
introduce two baselines: one is to activate all the links and the other is to
activate the links greedily to meet the TIN condition.

\begin{figure}[t]
\centering
\centerline{\includegraphics[width=9.5cm]{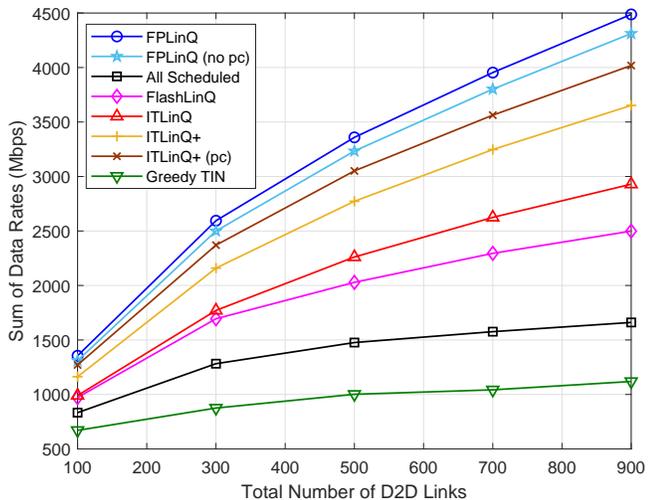}}
\caption{Sum-rate maximization for the single-association D2D network.}
\label{fig:sum_rate_disjoint}
\end{figure}

Fig.~\ref{fig:sum_rate_disjoint} shows the sum rate versus the total number
of D2D links. Observe that ITLinQ+ outperforms ITLinQ, and ITLinQ outperforms
FlashLinQ, as expected from the previous literature \cite{ITLinQ,ITLinQ+}. Without
power control, FPLinQ (no pc) significantly outperforms FlashLinQ, ITLinQ, and
ITLinQ+, especially when the D2D links are densely located in the area. In
particular, observe that Greedy TIN is even worse than simply scheduling all
the links because it is too conservative about the effect of interference. 
Further, as suggested in \cite{ITLinQ+}, we run
ITLinQ+ and the power control algorithm (e.g., the WMMSE method) alternatively
in order to account for joint scheduling and power control; this method is
referred to as ``ITLinQ (pc)''. However, the performance of ITLinQ+ with power
control is still inferior to that of FPLinQ and even that of FPLinQ (no pc).


\begin{figure}[t]
\centering
\centerline{\includegraphics[width=9.5cm]{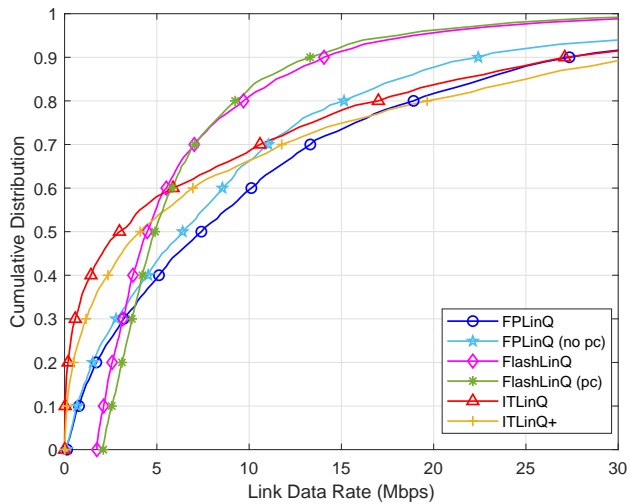}}
\caption{Log-utility maximization for the single-association D2D network.}
\label{fig:rcdf_disjoint}
\end{figure}

\begin{figure}[t]
\begin{minipage}[b]{1.0\linewidth}
\centering
\centerline{\includegraphics[width=9.6cm]{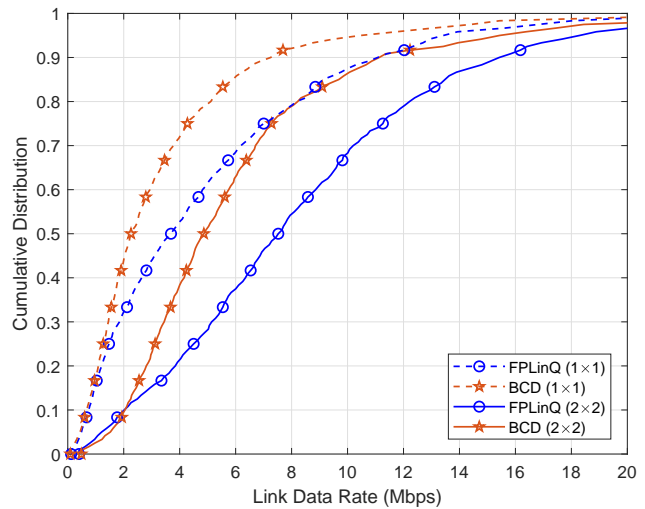}}
\caption{Log-utility maximization for the flexible-association D2D network: FPLinQ vs. BCD.}
\label{fig:rcdf_connected}
\end{minipage}
\end{figure}

The above simulation setting is only concerned with sum rate, as the weights are
all set to 1. We now consider a more demanding setting that takes priority
weights into account. In this simulation, the weights are updated using the
proportional fairness criterion, which is equivalent to maximizing the
log-utility of the average link rates in the long run
\cite{kushner}. The network setting follows the previous simulation; the total
number of links is fixed at 100. Fig.~\ref{fig:rcdf_disjoint} compares the
cumulative distribution of the link rates; the upper part of Table \ref{tab:utility}
compares the log-utility values. As we can see in Fig.~\ref{fig:rcdf_disjoint}, FPLinQ
(no pc) strikes a better balance between the high-rate regime and the low-rate
regime than ITLinQ and ITLinQ+. Surprisingly, FlashLinQ performs much better
than ITLinQ and ITLinQ+ in this simulation; its performance is even slightly better
than FPLinQ (no pc) according to Table \ref{tab:utility}. In particular,
observe in Fig.~\ref{fig:rcdf_disjoint} that the low-rate links benefit the
most from FlashLinQ, so FlashLinQ is fairly effective in protecting the
low-rate links from strong interference, but its threshold value must be chosen
carefully. Further, the benefit from the low-rate links comes at a cost for
high-rate links. Overall, when we include power control and compare FPLinQ
with a new benchmark method that combines FlashLinQ and power control in
an alternative fashion, referred to as ``FlashLinQ (pc)'',
FPLinQ outperforms FlashLinQ (pc) in network utility, when scheduling is
optimized along with transmit powers, as shown in Table \ref{tab:utility} on the next page.

\begin{figure}[t]
\begin{minipage}[b]{1.0\linewidth}
\centering
\centerline{\includegraphics[width=9.6cm]{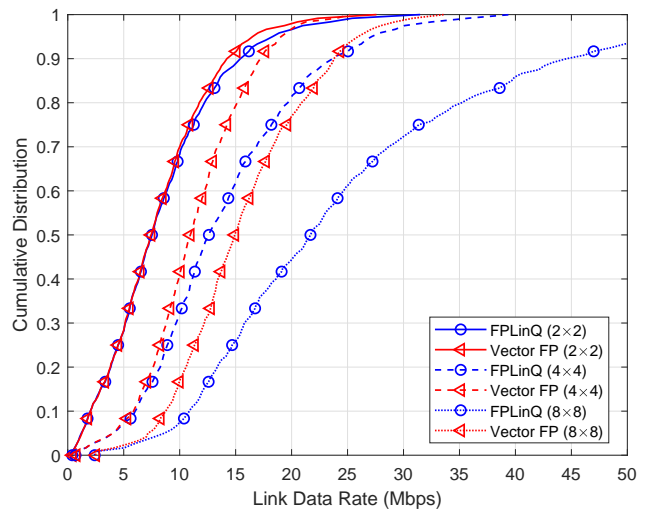}}
\caption{Log-utility maximization for the flexible-association D2D network: FPLinQ vs. Vector FP.}
\label{fig:rcdf_vectorFP}
\end{minipage}
\end{figure}

\begin{table}[t]
\small
\centering
\caption{\small Sum Log-Utility over D2D Networks}
\begin{tabular}{|c|c|}
\hline
Fixed Single Association &  Log Utility \\
\hline
FPLinQ  & 177.6 \\
FPLinQ (no pc) & 162.3 \\
FlashLinQ & 163.0 \\
FlashLinQ (pc) & 170.6 \\
ITLinQ & 57.0 \\
ITLInQ+ & 109.5 \\
\hline
\hline
Flexible Association &  Log Utility \\
\hline
BCD ($1\times1$)& 99.6\\
BCD ($2\times2$) & 186.4\\
\hline
FPLinQ $(1\times1)$ & 139.3\\
FPLinQ ($2\times2$) & 224.4\\
FPLinQ ($4\times4$) & 298.9 \\
FPLinQ ($8\times8$) & 369.0\\
\hline
Vector FP ($2\times2$) & 223.3\\
Vector FP ($4\times4$) & 279.0\\
Vector FP ($8\times8$) & 321.5\\
\hline
\end{tabular}
\label{tab:utility}
\end{table}

Finally, we consider the flexible association case. We first generate 100 disjoint
D2D links as before, but also generate two extra transmitters randomly for each
receiver, and further let one third of the transmitters connect with one additional
geographically closest receiver (excluding the already connected one). In
this setup, we frequently encounter the situation that multiple transmitters
contend for the same receiver, so the premature turning-off problem is very
likely to occur. We again optimize the log-utility by updating the link
weights according to the proportional fairness criterion. FPLinQ is compared
with the BCD method for both the single-antenna case and the $2\times 2$ MIMO
case (i.e., when each device terminal has $2$ antennas). Note that FlashLinQ, ITLinQ, and ITLinQ+ are not applicable here, because
they do not handle MIMO. Fig.~\ref{fig:rcdf_connected} shows the cumulative
distribution function of link rates, and the lower part of Table \ref{tab:utility}
summarizes the log-utility results. It can be seen that FPLinQ significantly
outperforms BCD. In fact, as shown in Fig.~\ref{fig:rcdf_connected}, FPLinQ
improves upon the BCD method by more than 50\% for the 50th percentile link rate,
in both the single-antenna case and the MIMO case. The corresponding
log-utility of FPLinQ is also much higher. These results show that the
premature turning-off can be fairly detrimental to the performance of D2D
system in the flexible association case, thus making the proposed FPLinQ strategy
a preferred strategy.

One of the key advantages of the proposed matrix FP strategy is its ability to accommodate multiple data streams in each MIMO link. In the next simulation, we evaluate the gain of multiple data-stream transmission over the single data-stream transmission. Toward this end, we compare FPLinQ (with matrix FP) against the vector FP method (also called multidimensional FP in \cite{part2}). The vector FP algorithm
is the same as Algorithm \ref{alg:FPLinQ} except that each transmit beamformer $\mathbf V_{i}\in\mathbb C^N$ is a complex vector instead of a matrix, so at most one data stream can be transmitted on each link. Fig.~\ref{fig:rcdf_vectorFP} shows the cumulative distribution function of link
rates under different MIMO settings. It can be seen that while the gain of FPLinQ as compared to the vector FP is marginal in the $2\times2$ MIMO case, as more antennas are deployed at each terminal, the multiple data-stream transmission by FPLinQ starts to significantly outperform. The above observations is also evident from the lower part of Table \ref{tab:utility}. Therefore, if the number of antenna $N$ is small (e.g., 2), then using the vector FP in Algorithm \ref{alg:FPLinQ} is more suited because of its lower complexity; on the other hand, if $N$ is large (e.g., 8), then using FPLinQ with multiple data-stream transmission can boost the overall network throughput significantly.

\begin{figure}[t]
\begin{minipage}[b]{1.0\linewidth}
\centering
\centerline{\includegraphics[width=9.6cm]{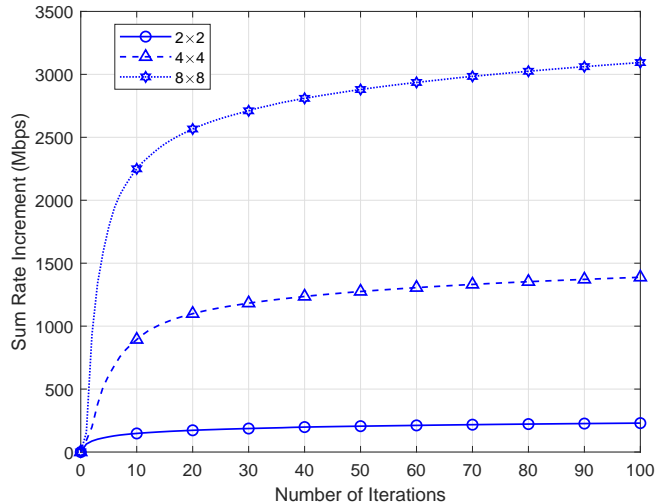}}
\caption{Convergence of FPLinQ in maximizing the sum rate for the flexible-association D2D network.}
\label{fig:convergence}
\end{minipage}
\end{figure}

Finally, Fig.~\ref{fig:convergence} shows the convergence speed of FPLinQ when
applied to maximizing the sum rate for the flexible-association D2D network with 400 links. Under the three MIMO settings (i.e., $2\times2$, $4\times4$, and $8\times8$), FPLinQ is observed to have fairly rapid convergence rate. Taking the $2\times2$ case for example, we observe from Fig.~\ref{fig:convergence} that the majority of sum rate increment is obtained after the first 10 iterations. We also see that the convergence of FPLinQ is slower when more antennas are deployed at each terminal. But, as shown in the $8\times8$ case in Fig.~\ref{fig:convergence}, we can already reap most of the rate gain after about 40-60 iterations.

\section{Conclusion}
\label{sec:conclusion}

This work proposes an interference-aware spectrum sharing strategy named FPLinQ
to coordinate the scheduling decisions along with beamforming and power control
across the wireless D2D links. The key step is to treat the weighted sum-rate
maximization as a matrix FP problem and to use a sequence of matrix FP
transforms to allow iterative optimization of scheduling and beamforming. We
show that FPLinQ is closely related to the MM algorithm, thus its convergence
is guaranteed. As compared to the existing methods, FPLinQ does not involve
tuning of design parameters and does not suffer from the premature turning-off
problem.  The numerical results show that FPLinQ outperforms the
state-of-the-art methods in terms of sum-rate maximization and log-utility
maximization.

%

\bibliographystyle{IEEEbib}
\bibliography{IEEEabrv,strings}

\end{document}